\documentclass[twocolumn,showpacs,superscriptaddress,nofootinbib,floatfix]{revtex4}
\usepackage{graphicx}
\begin{document}
\def\Journal#1#2#3#4{{#1} {\bf #2}, #3 (#4)}
\title{\bf A Comparison between a Minijet Model and a Glasma Flux Tube Model for Central Au-Au Collisions at $\sqrt{s_{NN}}$=200 GeV.}
\author{R.S.~Longacre}\affiliation{Brookhaven National Laboratory, Upton, New York 11973}
\date{\today}
 
\begin{abstract}
In is paper we compare two models with central Au-Au collisions at
$\sqrt{s_{NN}}$=200 GeV. The first model is a minijet model which assumes that
around $\sim$50 minijets are produced in back-to-back pairs and have an altered
fragmentation functions. It is also assumed that the fragments are transparent
and escape the collision zone and are detected. The second model is a glasma
flux tube model which leads to flux tubes on the surface of a radial expanding
fireball driven by interacting flux tubes near the center of the fireball 
through plasma instabilities. This internal fireball becomes an opaque
hydro fluid which pushes the surface flux tubes outward. Around $\sim$12 
surface flux tubes remain and fragment with $\sim$1/2 the produced particles
escaping the collision zone and are detected. Both models can reproduce two
particle angular correlations in the different $p_{t1}$ $p_{t2}$ bins. We
also compare the two models for three additional effects: meson baryon ratios;
the long range nearside correlation called the ridge; and the so-called mach
cone effect when applied to three particle angular correlations.
\end{abstract}

\pacs{25.75.Nq, 11.30.Er, 25.75.Gz, 12.38.Mh}
 
\maketitle

\section{Introduction and review of models} 

In this paper we discuss two models. The first model is a minijet 
model\cite{Trainor1}. The second is a glasma flux tube model 
(GFTM)\cite{Dumitru}. 

The paper is organized in the following manner:

Sec. 1 is the introduction and review of models. Sec. 2 discuss two particle
angular correlation in the two models. Sec. 3 discuss baryon and
anti-baryon formation in both models. Sec. 4 demonstrates how the
ridge is formed by flux tubes when a jet trigger is added to the GFTM. 
Sec. 5 treats the so-called mach cone effect by analyzing three particle 
angular correlations in the two models. Sec. 6 presents the summary and 
discussion.

\subsection{Minijet Model} 

    The analysis of angular correlations led to unanticipated structure in the
final state of p-p and Au-Au collisions, subsequently identified with parton 
fragmentation in the form of minijets[3-7]. Two-component analysis of p-p and
Au-Au spectra revealed a corresponding hard component, a minimum-bias fragment
distribution associated with minijets, suggesting that jet phenomena extend 
down to 0.1 GeV/c\cite{Trainor2,star4}. From a given p-p or Au-Au collisions
particles are produced with three kinematic variables ($\eta$,$\phi$,$p_t$).
The $p_t$ spectra provide information about parton fragmentation, but the 
fragmentation process is more accessible on a logarithmic momentum variable.
The dominant hadron (pion) is ultrarelativistic at 1 GeV/c, and a relativistic 
kinematic variable is a reasonable alternative. Transverse rapidity in a 
longitudinally comoving frame near midrapidity $\eta$=0 is defined by  

\begin{equation}
y_t = ln([m_t + p_t]/m_0),
\end{equation}

with $m_0$ a hadron mass and $m_t^2$ = $p_t^2$ + $m_0^2$. If one integrates
over $\phi$ the event multiplicity ($dn/d\eta$ $\equiv$ $\rho$, the 1D density
on $\eta$) can be written for p-p collision in a two component model as

\begin{equation}
\rho(y_t;\eta=0)=S_0(y_t)+H_0(y_t).
\end{equation}

$S_0$ is a Levy distribution on $m_t$ which represents soft processes and $H_0$
is a Gaussian on $y_t$ which represents hard processes. The soft process has 
no $\phi$ dependence but does have a Gaussian correlation in the longitudinal 
direction($\eta$). This correlation can be expressed in a two particle way as
$\Delta \eta = \eta_1 - \eta_2$ which is the difference of the psuedorapidity.

\begin{equation}
\rho_s(\Delta \eta)=A_0exp[-\Delta \eta^2/2\sigma_0^2].
\end{equation}

The hard process arise from a scattering of two partons thus minijets are 
formed. Each minijet fragments along its parton axis and generates a 2D
correlation $\Delta \phi = \phi_1 - \phi_2$ and $\Delta \eta = 
\eta_1 - \eta_2$. 

\begin{equation}
\rho_h(\Delta \phi,\Delta \eta)=A_hexp[-\Delta \phi^2/2\sigma_{\phi}^2]exp[-\Delta \eta^2/2\sigma_{\eta}^2].
\end{equation}

Since for every minijet fragmenting leading to a peak at $\Delta \phi$ = 
$0^\circ$, there is its scattered partner in the backward direction 
$\Delta \phi$ = $180^\circ$. The backward scattered minijet will range over
many psuedorapidity values, therefore its correlation with the fragmentation 
of the near side minijet will have a broad $\Delta \eta$ width. In this
condition simple momentum conservation is good enough and that is a 
$-cos(\Delta \phi)$ term.

The two-component model of hadron production in A-A collisions assumes that the
soft component is proportional to the participant pair number (linear 
superposition of N-N collisions), and the hard component is proportional to
the number of N-N binary collisions (parton scattering)\cite{Nardi}. Any
deviations from the model are specific to A-A collisions and may reveal 
properties of an A-A medium. In terms of mean participant path length
$\nu$ = 2$n_{bin}$/$n_{part}$ the $p_t$-integrated A-A hadron density on 
$\eta$ is

\begin{equation}
{2 \over n_{part}} {dn \over d\eta_{AA}} = \rho_s + \nu \rho_h .
\end{equation}

By analogy with Eg. (2) A-A density as a function of centrality parameter $\nu$
becomes

\begin{equation}
{2 \over n_{part}} \rho_{AA}(y_t;\nu,\eta=0) = S_{NN}(y_t) + \nu r_{AA}(y_t;\nu) H_{NN}(y_t) .
\end{equation}

In the above equation $S_{NN}(y_t)$ = $S_0(y_t)$ and $H_{NN}(y_t)$ = $H_0(y_t)$
which is p-p scattering or nucleon-nucleon scattering. We also can define for 
the hard density for A-A collisions in terms of $\nu$ as 
$H_{AA}(y_t;\nu)$ = $r_{AA}(y_t;\nu) H_{NN}(y_t)$. The density 2/$n_{part}$ 
1/2$\pi$ 1/$y_t$ $d^2$/$dy_t$$d\eta$ for pions as a function of $y_t$ at 
$\sqrt{s_{NN}}$=200 GeV for p-p, Au-Au, and $H_{NN}(y_t)$ is shown in Figure 1.

\begin{figure*}[ht] \centerline{\includegraphics[width=0.800\textwidth]
{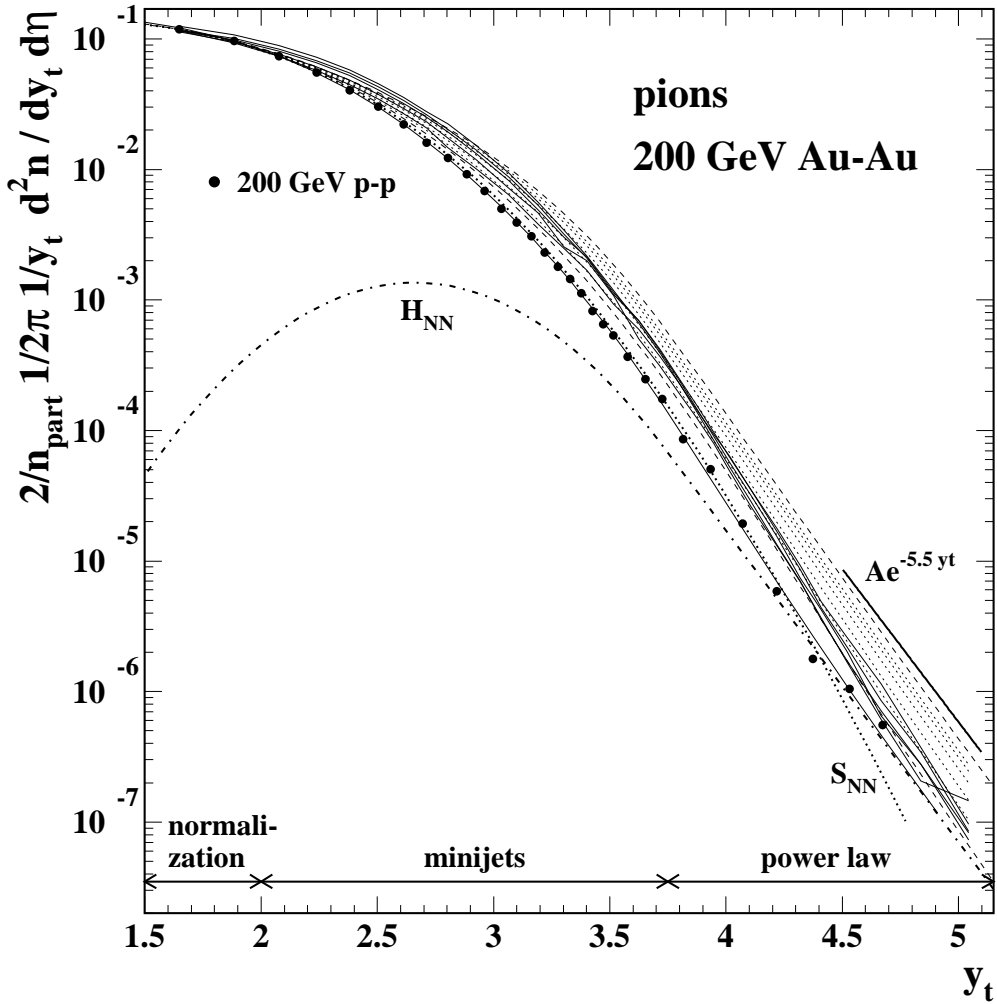}} \caption{Pion $y_t$ density for five Au-Au centralities 
(solid curves). Density for p-p is also shown (solid dots). $H_{NN}$ is the 
hard Gaussian on $y_t$ which represents hard processes (dash-dot curve).}
\label{fig1}
\end{figure*}

We see from the above equation the $S_{NN}(y_t)$ is universal and scales with
the participant pairs. This means we can extract $\nu H_{AA}(y_t)$ from the 
density measured in p-p and Au-Au at $\sqrt{s_{NN}}$=200 GeV from Figure 1 
giving us Figure 2. 

\begin{figure*}[ht] \centerline{\includegraphics[width=0.800\textwidth]
{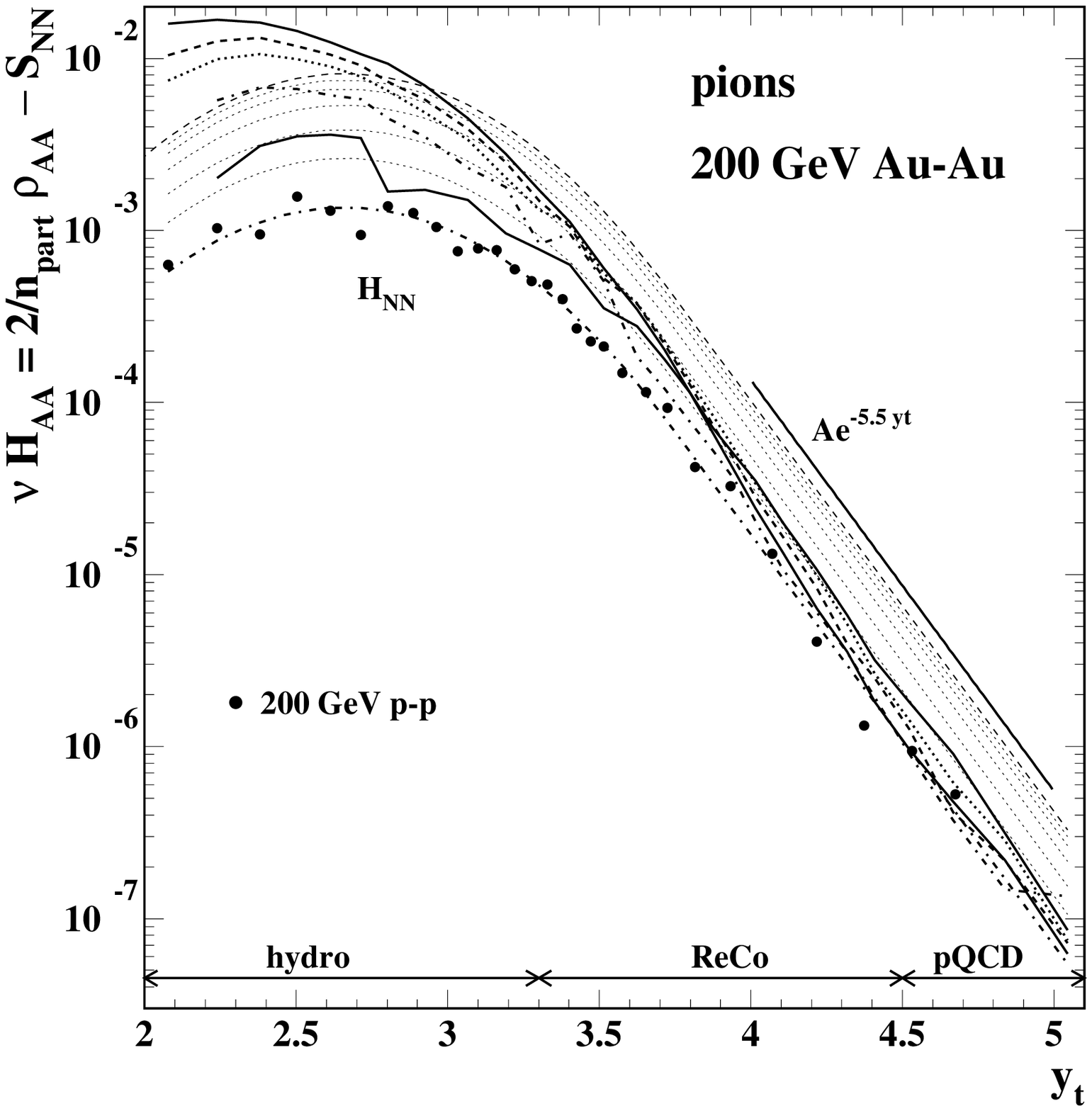}} \caption{The hard component of pion $y_t$ spectra in the 
form $\nu H_{AA}(y_t)$ (thicker curves with changing line style) compared 
to two-component reference $\nu H_{NN}(y_t)$ (dotted curve). The dashed 
reference curves are limiting cases for $\nu$ = 1, 6.} 
\label{fig2}
\end{figure*}
 
Finally the ratio $H_{AA}$/$H_{NN}$ ($R_{AA}$) is plotted in Figure 3. In the
Au-Au central collision (0-12\%) at a $y_t$ value of 2 ($p_t$=0.5 GeV/c) there 
is 5 times as many pions coming from minijet fragmentation as there is in a N-N
collision. This implies a large increase in correlated pion fragments and 
should show up as an increased in two particle angular correlations. We will
see this in Sec. 2 where we show these angular correlations and discuss the 
number of particles in the minijets.

\begin{figure*}[ht] \centerline{\includegraphics[width=0.800\textwidth]
{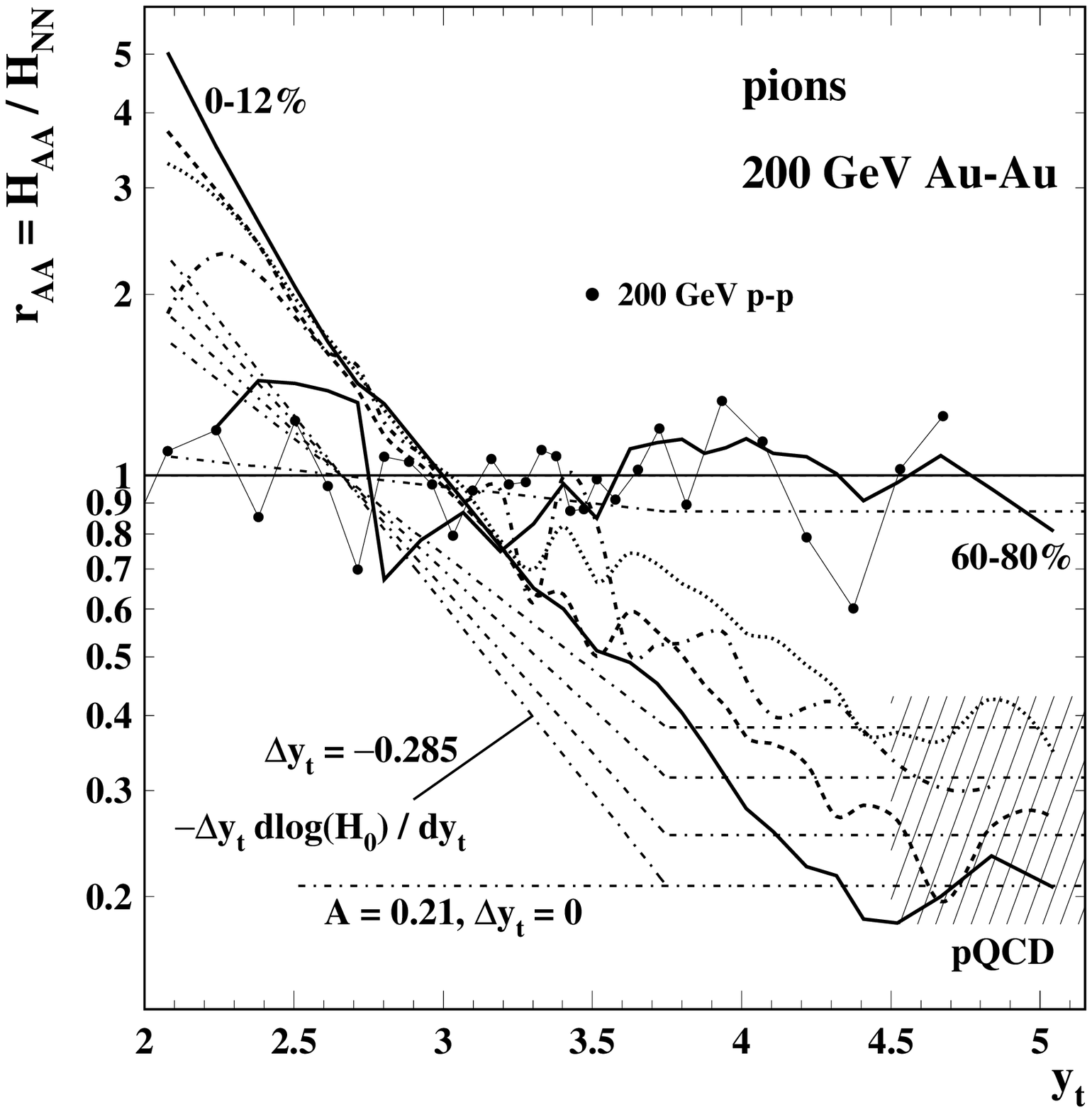}} \caption{Hard-component ratio for pions and five Au-Au 
centralities (thicker curves with changing line styles) relative to the 
N-N hard-component reference. The connected dots are data from NSD p-p 
collisions.}
\label{fig3}
\end{figure*}

The common measurement of parton energy loss is expressed as a nuclear 
modification factor $R_{AA}$. The hard-component ratio $r_{AA}$ measured over
$y_t$ ($p_t$) provides similar information. Note that $y_t$ = 5 
($p_t$=10 GeV/c) $r_{AA}$ = 0.2 which is the same value calculated for 
$R_{AA}$. This value is considered caused by jet quenching as partons are
absorbed by the opaque medium. However in the minijet picture suggests that
no partons are lost in A-A collisions. Their manifestation (in spectrum
structure and correlations) is simply redistributed within the fragment
momentum distribution, and the fragment number increases. A high-$p_t$ 
triggered jet yield may be reduced by a factor of five within a particular 
$p_t$ range, but additional fragments emerge elsewhere, still with jet-like 
correlation structure\cite{star1,star2,Kettler}.

\subsection{Glasma Flux Tube Model} 

A glasma flux tube model (GFTM)\cite{Dumitru} that had been developed considers
that the wavefunctions of the incoming projectiles, form sheets of color glass 
condensates (CGC)\cite{CGC} that at high energies collide, interact, and 
evolve into high intensity color electric and magnetic fields. This collection 
of primordial fields is the Glasma\cite{Lappi,Gelis}, and initially it is 
composed of only rapidity independent longitudinal color electric and magnetic 
fields. An essential feature of the Glasma is that the fields are localized 
in the transverse space of the collision zone with a size of 1/$Q_s$. $Q_s$ is
the saturation momentum of partons in the nuclear wavefunction. These
longitudinal color electric and magnetic fields generate topological 
Chern-Simons charge\cite{Simons} which becomes a source for particle 
production.

The transverse space is filled with flux tubes of large longitudinal extent 
but small transverse size $\sim$$Q^{-1}_s$. Particle production from a flux 
tube is a Poisson process, since the flux tube is a coherent state. 
The flux tubes at the center of the transverse plane interact with each other
through plasma instabilities\cite{Lappi,Romatschke1} and create a locally 
thermalized system, where partons emitted from these flux tubes locally 
equilibrate. A hydro system with transverse flow builds causing a radially
flowing blast wave\cite{Gavin}. The flux tubes that are near the surface of the
fireball get the largest radial flow and are emitted from the surface.
 
$Q_s$ is around 1 GeV/c thus the transverse size of the flux tube is about 
1/4fm. The flux tubes near the surface are initially at a radius $\sim$5fm. 
The $\phi$ angle wedge of the flux tube is $\sim$1/20 radians or 
$\sim$$3^\circ$. Thus the flux tube initially has a narrow range in $\phi$. The
width in $\Delta \eta$ correlation of particles results from the 
independent longitudinal color electric and magnetic fields that created the 
Glasma flux tubes. In this paper we relate particle production from the 
surface flux tube to a related model Parton Bubble Model(PBM)\cite{PBM}. It was
shown in Ref.\cite{PBMGFTM} that for central Au-Au collisions at 
$\sqrt{s_{NN}}$=200 the PBM is a good approximation to the GFTM surface flux 
tube formation. 

The flux tubes on the surface turn out to be on the average 12 in number. They
form an approximate ring about the center of the collision see Figure 4. The 
twelve tube ring creates the average behavior of tube survival near the 
surface of the expanding fire ball of the blast wave. The final state surface 
tubes that emit the final state particles at kinetic freezeout are given
by the PBM. One should note that the blast wave surface is moving at its 
maximum velocity at freezeout (3c/4). 

\begin{figure*}[ht] \centerline{\includegraphics[width=0.800\textwidth]
{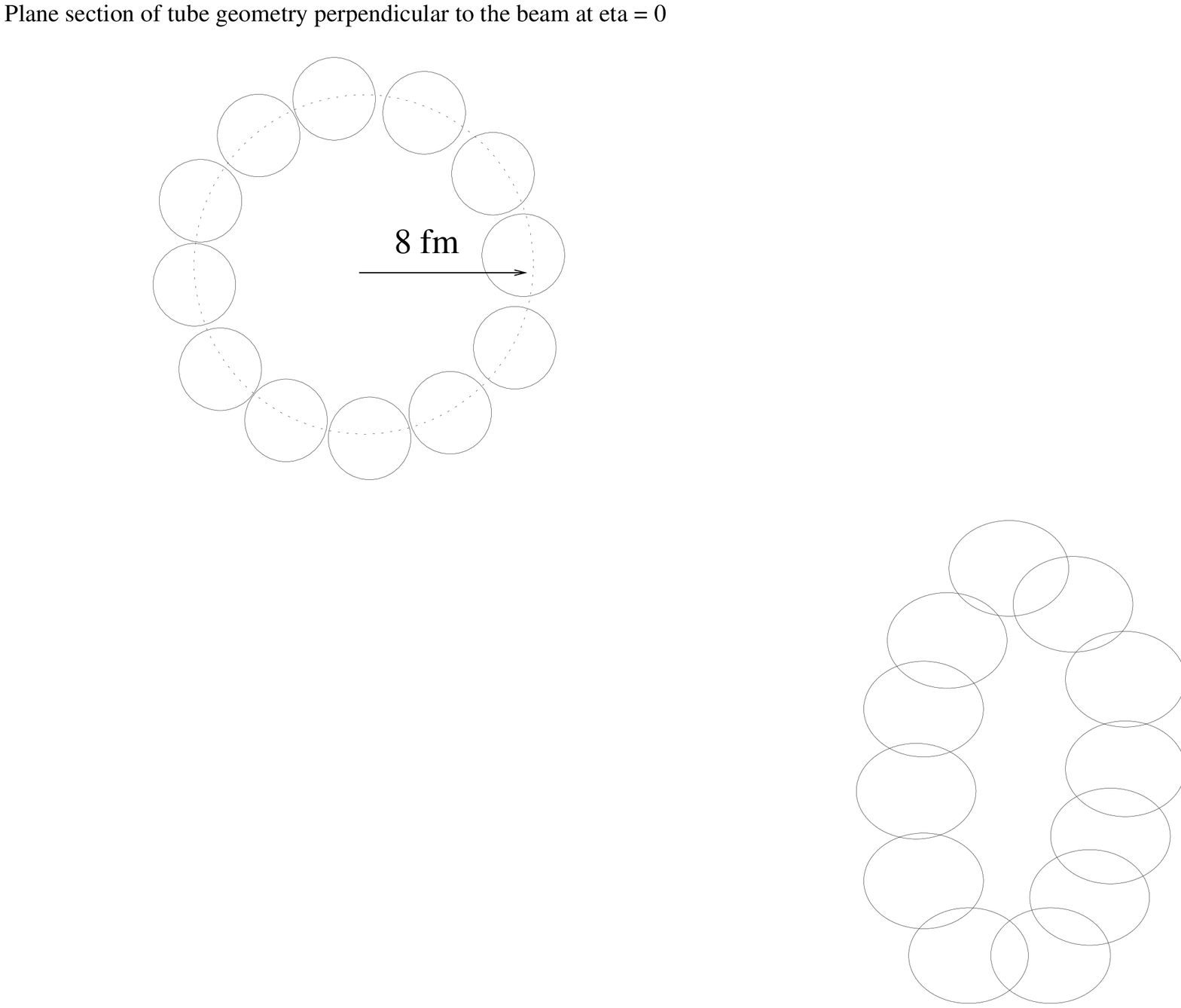}} \caption{The tube geometry is an 8 fm radius ring perpendicular 
to and centered on the beam axis. It is composed of twelve adjacent 2 fm radius
circular tubes elongated along the beam direction as part of the flux
tube geometry. We project on a plane section perpendicular to the beam axis.}
\label{fig4}
\end{figure*}

 The space momentum correlation of the blast wave provides us with a strong 
angular correlation signal. PYTHIA fragmentation functions\cite{pythia} were 
used for the tube fragmentation that generate the final state particles 
emitted from the tube. The initial transverse size of a flux tube 
$\sim$1/4fm has expanded to the size of $\sim$2fm at kinetic freezeout. 
Many particles that come from the surface of the fireball will have a 
$p_t$ greater than 0.8 Gev/c. The final state tube size and the Hanbury-Brown 
and Twiss (HBT) observations\cite{HBT} of pions that have a momentum range 
greater than 0.8 GeV/c are consistent both being $\sim$2fm. A single parton 
using PYTHIA forms a jet with the parton having a fixed $\eta$ and $\phi$ 
(see Figure 5). For central events each of the twelve tubes have 3-4 partons 
per tube each at a fixed $\phi$ for a given tube. The $p_t$ distribution of 
the partons is similar to pQCD but has a suppression at high $p_t$ like the 
data. The 3-4 partons in the tube which shower using PYTHIA all have a 
different $\eta$ values but all have the same $\phi$ (see Figure 6). The 
PBM explained the high precision Au-Au central (0-10\%) collisions at 
$\sqrt{s_{NN}} =$ 200 GeV\cite{centralproduction}(the highest RHIC energy).

\begin{figure*}[ht] \centerline{\includegraphics[width=0.800\textwidth]
{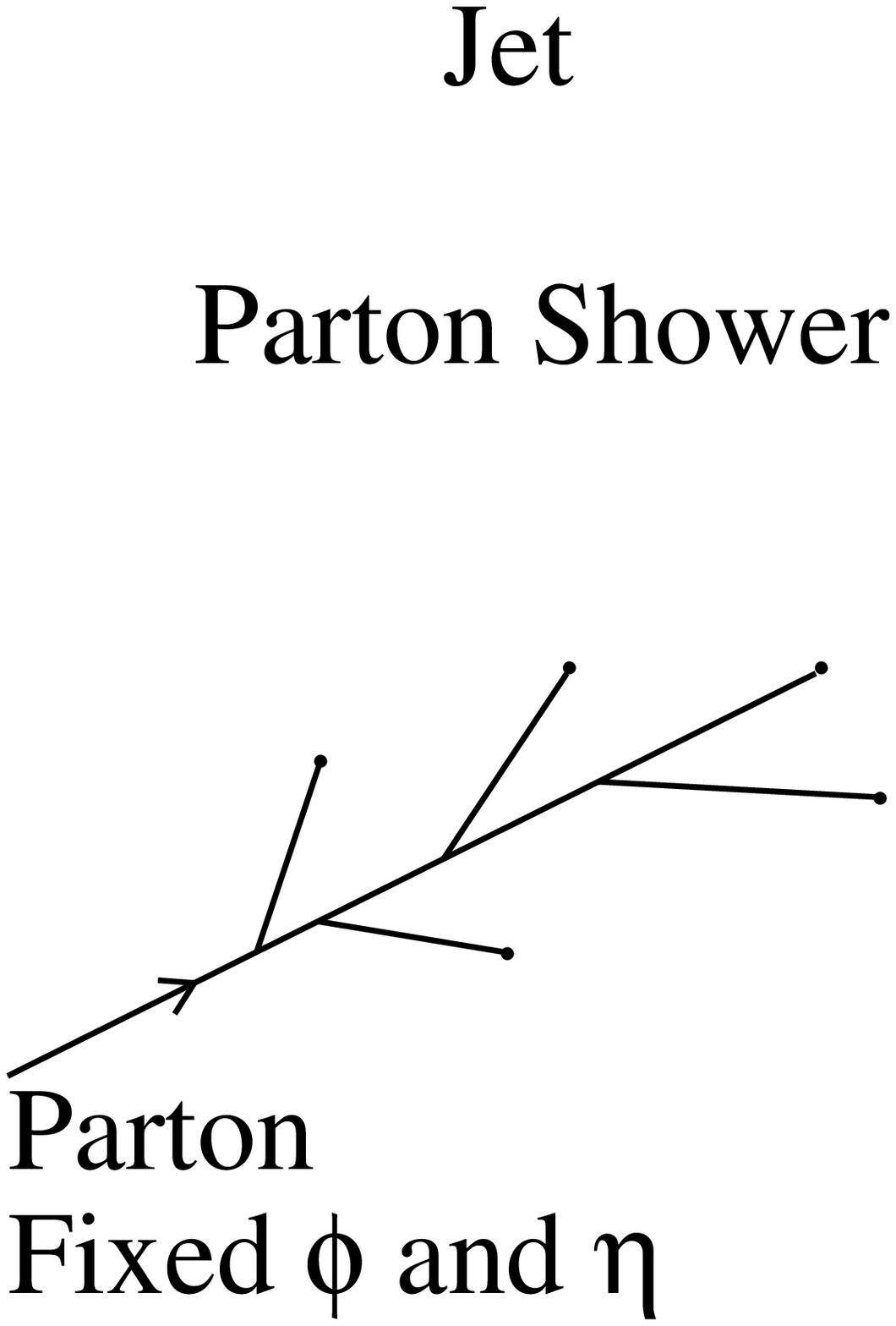}} \caption{A jet parton shower.}
\label{fig5}
\end{figure*}

\begin{figure*}[ht] \centerline{\includegraphics[width=0.800\textwidth]
{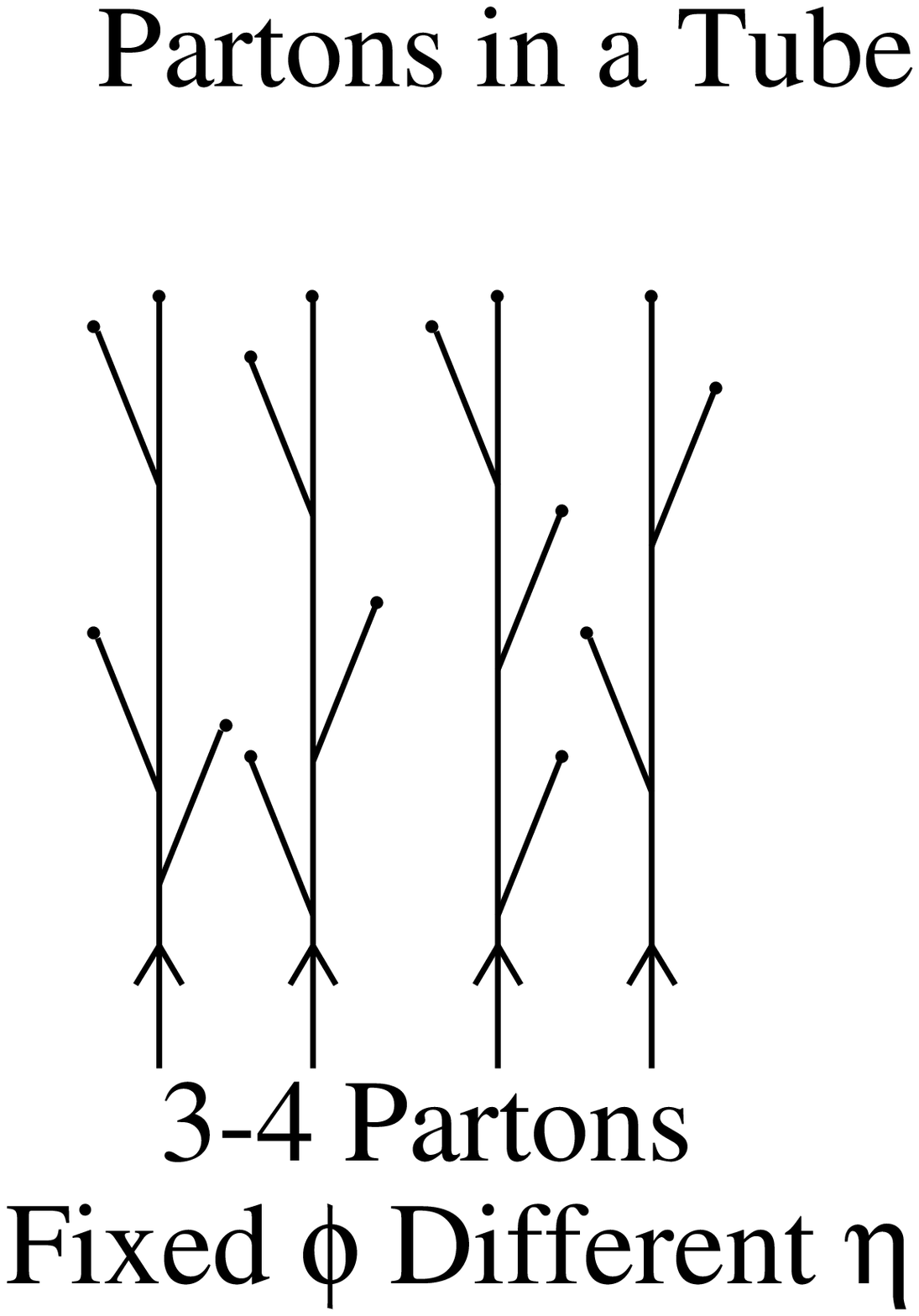}} \caption{Each tube contains 3-4 partons as shown.}
\label{fig6}
\end{figure*}

\section{The Correlation Function for Central Au-Au Data}

We utilize a two particle correlation function in the two dimensional (2-D)
space\footnote{$\Delta \phi = \phi_1 - \phi_2$ where $\phi$ is the azimuthal
angle of a particle measured in a clockwise direction about the beam.
$\Delta \eta = \eta_1 - \eta_2$ which is the difference of the psuedorapidity
of the pair of particles} of $\Delta \phi$ versus $\Delta \eta$. The 2-D 
total correlation function is defined as:

\begin{equation}
C(\Delta \phi,\Delta \eta)
=S(\Delta \phi,\Delta \eta)/M(\Delta \phi,\Delta \eta).
\end{equation}

Where S($\Delta \phi,\Delta \eta$) is the number of pairs at the corresponding
values of $\Delta \phi,\Delta \eta$ coming from the same event, after we have
summed over all the events. M($\Delta \phi,\Delta \eta$) is the number of pairs
at the corresponding values of $\Delta \phi,\Delta \eta$ coming from the mixed
events, after we have summed over all our created mixed events. A mixed event
pair has each of the two particles chosen from a different event. We make on
the order of ten times the number of mixed events as real events. We rescale
the number of pairs in the mixed events to be equal to the number of pairs in
the real events. This procedure implies a binning in order to deal with finite
statistics. The division by M($\Delta \phi,\Delta \eta$) for experimental data
essentially removes or drastically reduces acceptance and instrumental effects.
If the mixed pair distribution was the same as the real pair distribution 
C($\Delta \phi,\Delta \eta$) should have unit value for all of the binned 
$\Delta \phi,\Delta \eta$. In the correlations used in this paper we select 
particles independent of its charge. The correlation of this type is called 
a Charge Independent (CI) Correlation. This difference correlation function 
has the defined property that it only depends on the differences of the 
azimuthal angle ($\Delta \phi$) and the beam angle ($\Delta \eta$) for the 
two particle pair. Thus the two dimensional difference correlation distribution
for each tube or minijet which is part of C($\Delta \phi,\Delta \eta$) is 
similar for each of the objects and will image on top of each other. We
further divide the data (see Table I) into $p_t$ ranges (bins). 

\begin{center}
\begin{table}
\begin{tabular}{ c r }\hline\hline
$p_t$ range & amount \\ \hline
$4.0 GeV/c - 1.1 GeV/c$ & 149 \\ \hline
$1.1 GeV/c - 0.8 GeV/c$ & 171 \\ \hline
$0.8 GeV/c - 0.65 GeV/c$ & 152 \\ \hline
$0.65 GeV/c - 0.5 GeV/c$ & 230 \\ \hline
$0.5 GeV/c - 0.4 GeV/c$ & 208 \\ \hline
$0.4 GeV/c - 0.3 GeV/c$ & 260 \\ \hline
$0.3 GeV/c - 0.2 GeV/c$ & 291 \\ \hline
\end{tabular}
\caption[]{The $p_t$ bins and the number of charged particles per bin 
with $| \eta |$ $<$ 1.0.}
\end{table}
\end{center}

Since we are choosing particle pairs, we choose for the first particle $p_{t1}$
which could be in one bin and for the second particle $p_{t2}$ which could be 
in another bin. Therefore binning implies a matrix of $p_{t1}$ vs $p_{t2}$. We 
have have 7 bins thus there are 28 independent combinations. Each of the 
combinations will have a different number of enters. In order to take out this 
difference one uses multiplicity scaling\cite{PBME,centralitydependence}. The 
diagonal bins one scales event average of Table I. For the off diagonal 
combinations one uses the product of square root of corresponding diagonal 
event averages. In Figure 7 we show the correlation function equation 7 for the
highest diagonal bin $p_t$ 4.0 to 1.1 GeV/c. Figure 8 is the smallest diagonal 
bin $p_t$ 0.3 to 0.2 GeV/c.

\begin{figure*}[ht] \centerline{\includegraphics[width=0.800\textwidth]
{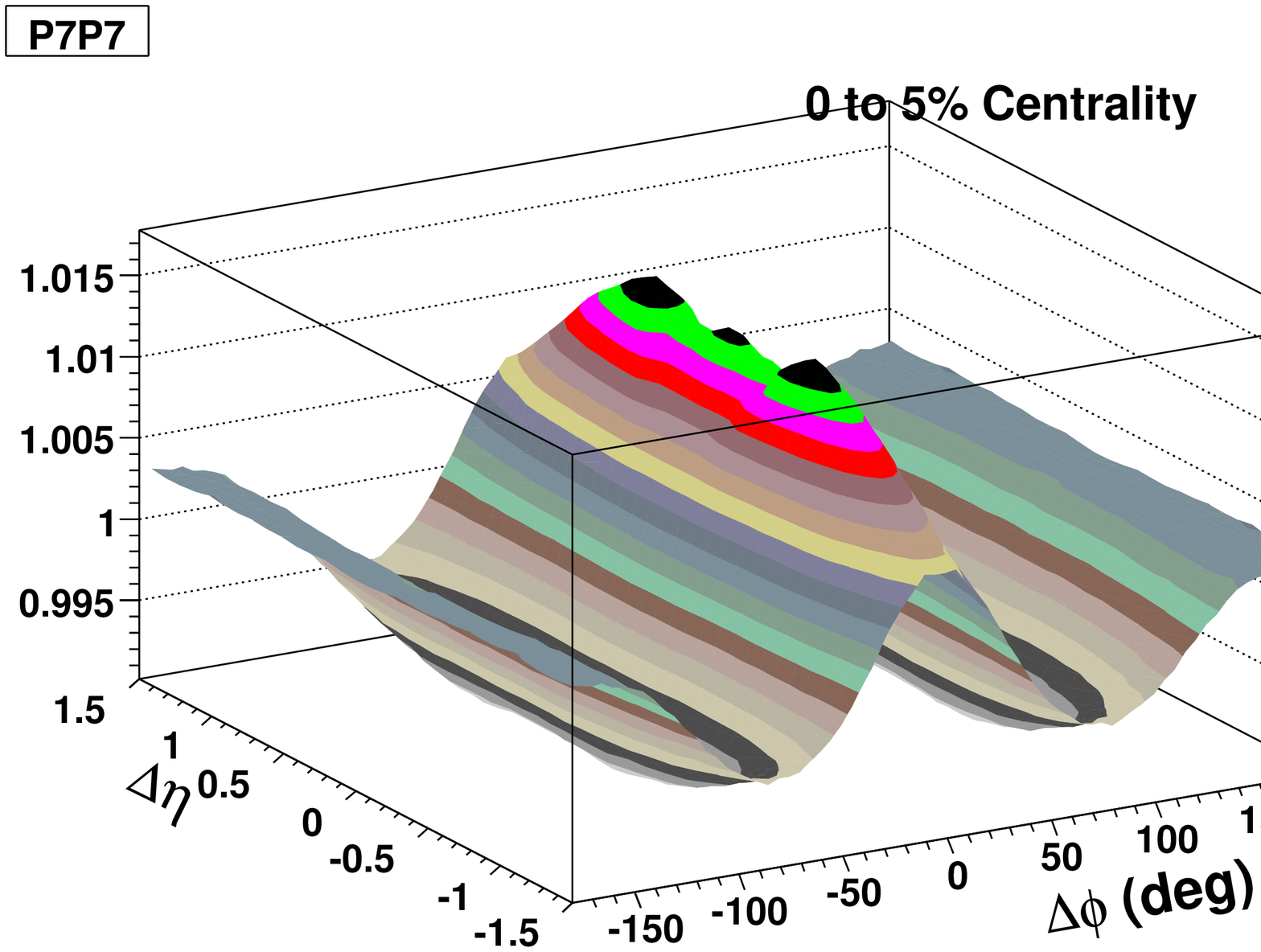}} \caption{$\Delta \phi$ vs.$\Delta \eta$ CI correlation for the 
0-5\% centrality bin for Au-Au collisions at $\sqrt{s_{NN}} = $ 200 GeV 
requiring both particles to be in bin 7 $p_t$ greater than 1.1 GeV/c and $p_t$ 
less than 4.0 GeV/c.}
\label{fig7}
\end{figure*}

\begin{figure*}[ht] \centerline{\includegraphics[width=0.800\textwidth]
{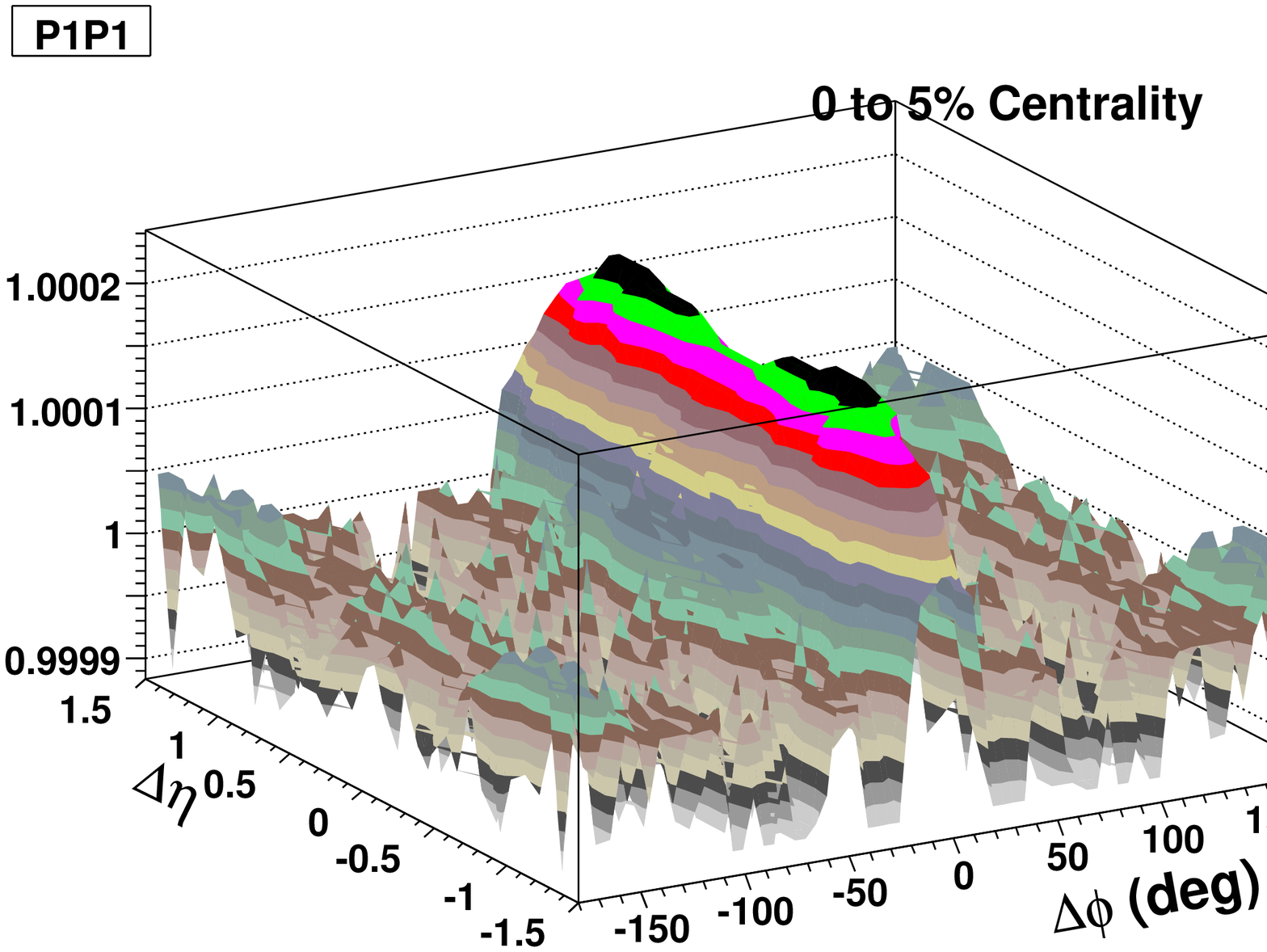}} \caption{$\Delta \phi$ vs.$\Delta \eta$ CI correlation for 
the 0-5\% centrality bin for Au-Au collisions at $\sqrt{s_{NN}} = $ 200 GeV 
requiring both particles to be in bin 1 $p_t$ greater than 0.2 GeV/cand $p_t$ 
less than 0.3 GeV/c.}
\label{fig8}
\end{figure*}

Once we use multiplicity scaling we can compare all 28 combinations. 
In Figure 9 we show the 28 plots all having
the same scale. This make it easy to see how fast the correlation signals
drop off with lowering the momentum. These plots show the properties of
parton fragmentation. P1P7 has the same signal as P2P6, P3P5, and P4P4.
It should be noted that both the minijet model and GFTM give the same two
particle correlations.

\begin{figure*}[ht] \centerline{\includegraphics[width=0.800\textwidth]
{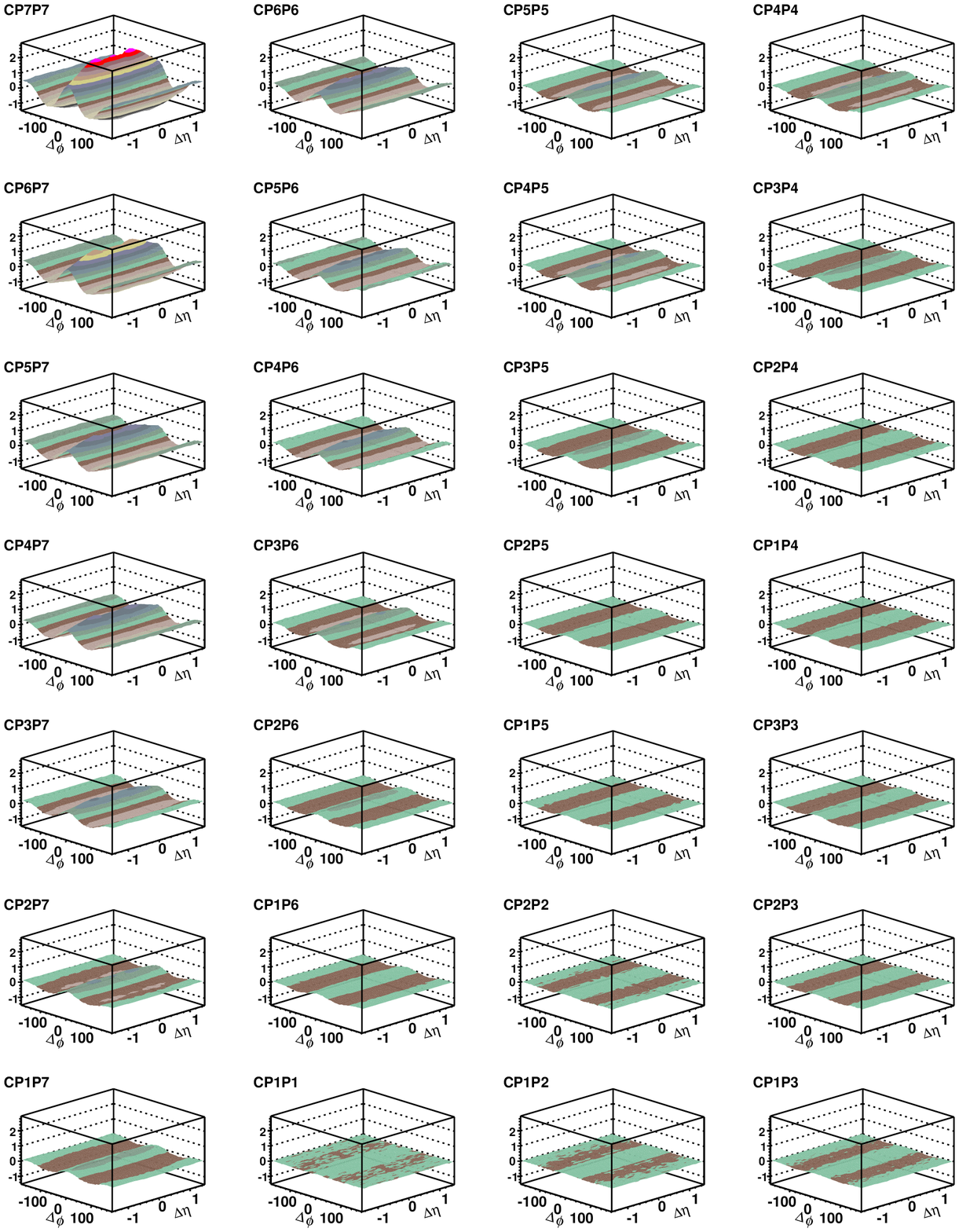}} \caption{$\Delta \phi$ vs.$\Delta \eta$ CI correlation for the 
0-5\% centrality bin for Au-Au collisions at $\sqrt{s_{NN}} = $ 200 GeV for 
all 28 combination of $p_{t1}$ vs $p_{t2}$ not rescaled (see text).}
\label{fig9}
\end{figure*}

\subsection{The properties of the minijet model}

For the above correlations on average there was 48 minijets per central
Au-Au collision. Each minijet on the average showered into 13 charged
particles. The soft uncorrelated particles accounted for 837 or 57\%
of the charged particles. This means that 43\% of the charged particles
come from minijet fragmentation (see Table II). All of the particles
coming from minijet fragmentation add toward the final observed correlation
signal with none being absorbed. The fact that the spectrum has been soften
and spread out in the beam direction ($\eta$), is a medium modification which
has not been yet calculated using QCD.

\begin{center}
\begin{table}
\begin{tabular}{ c r r }\hline\hline
variable & amount & fluctuations \\ \hline
$minijets$ & 48 & 4  \\ \hline
$particles$& 13  & 4  \\ \hline
$soft$ & 837 & 29  \\ \hline
\end{tabular}
\caption[]{Parameters of the minijet model for charged particles.}
\end{table}
\end{center}

\subsection{The properties of the Glasma Flux Tube Model}

For the above correlations on average there was 12 final state tubes on
the surface of the fireball per central Au-Au collision. Each tube on the 
average showered into 49 charged particles. The soft uncorrelated particles 
accounted for 873 or 60\% of the charged particles. Since the tubes are
sitting on the surface of the fireball and being push outward by radial flow,
not all particles emitted from the tube will escape. Approximately one half of
the particles that are on the outward surface leave the fireball and the other 
half are absorbed by the fireball (see Figure 10). This means that 20\% of 
the charged particles come from tube emission, and 294 particles are added to 
the soft particles increasing the number to 1167 (see Table III).

\begin{figure*}[ht] \centerline{\includegraphics[width=0.800\textwidth]
{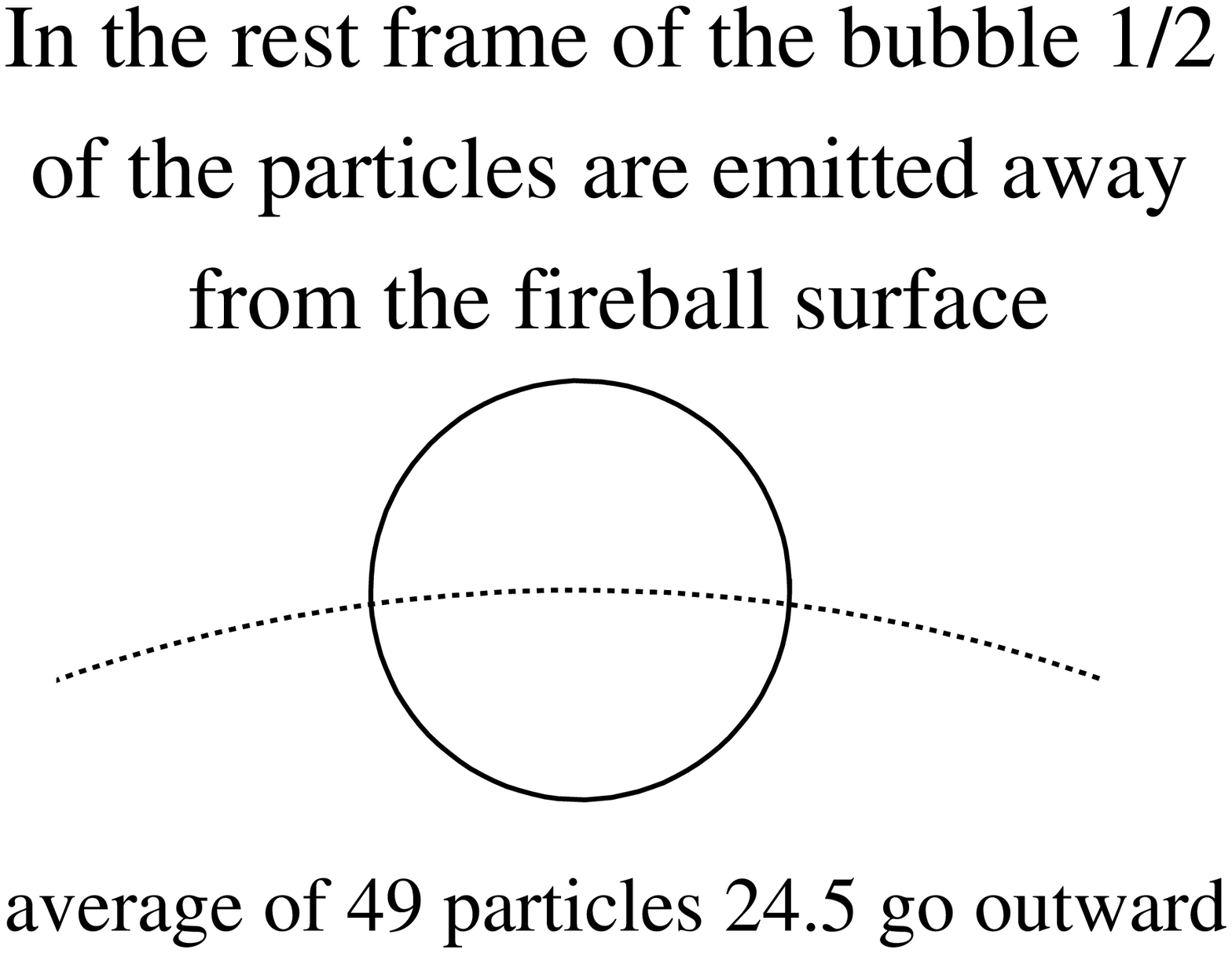}} \caption{Since the tubes are sitting on the surface of the 
fireball and being push outward by radial flow, not all particles emitted from 
the tube will escape. Approximately one half of the particles that are on the 
outward surface leave the fireball and the other half are absorbed by the 
fireball.} 
\label{fig10}
\end{figure*}

\begin{center}
\begin{table}
\begin{tabular}{ c r r }\hline\hline
variable & amount & fluctuations \\ \hline
$tubes$ & 12 & 0  \\ \hline
$particles$& 24.5  & 5  \\ \hline
$soft$ & 1167 & 34  \\ \hline
\end{tabular}
\caption[]{Parameters of the GFTM for charged particles.}
\end{table}
\end{center}

The particles that emitted outward are boosted in momentum, while the inward
particles are absorbed by the fireball. Out of the initial 49 particles per
tube the lower $p_t$ particles have larger losses. In Table IV we give a
detailed account of these percentage losses and give the average number of
charged particles coming from each tube for each $p_t$ bin.

\begin{center}
\begin{table}
\begin{tabular}{ c r r }\hline\hline
$p_t$ & amount & \%survive \\ \hline
$4.0 GeV/c - 1.1 GeV/c$ & 4.2 & 100 \\ \hline
$1.1 GeV/c - 0.8 GeV/c$ & 3.8 & 76 \\ \hline
$0.8 GeV/c - 0.65 GeV/c$ & 3.2 & 65 \\ \hline
$0.65 GeV/c - 0.5 GeV/c$ & 4.2 & 54 \\ \hline
$0.5 GeV/c - 0.4 GeV/c$ & 3.2 & 43 \\ \hline
$0.4 GeV/c - 0.3 GeV/c$ & 3.3 & 35 \\ \hline
$0.3 GeV/c - 0.2 GeV/c$ & 2.6 & 25 \\ \hline
\end{tabular}
\caption[]{Parameters of the GFTM for $p_t$ of the charged particles.}
\end{table}
\end{center}
 
In the surface GFTM we have thermalization and hydro flow for the soft 
particles, while all the two particle angular correlations come from
the tubes on the surface. The charge particle spectrum of the GFTM is given
by a blastwave model and the direct tube fragmentation is only 20\% of this
spectrum. The initial anisotropic azimuthal distribution of flux tubes is 
transported to the final state leaving its pattern on the ring of 
final state flux tubes on the surface. This final state anisotropic flow
pattern can be decomposed in a Fourier series ($v_1$, $v_2$, $v_3$, ...).
These coefficients have been measure\cite{Alver} and have been found to be
important in central Au-Au collisions. We will come back to these terms later
on when we consider three particle angular correlations. 

\section{Formation of baryon and anti-baryon in both models}

It was shown above that charged particle production differ between p-p and 
Au-Au at $\sqrt{s_{NN}} =$ 200. High $p_t$ charged particles in Au-Au are 
suppressed compared to p-p for $p_t$ $>$ 2.0 GeV/c ($y_t$ $>$ 3.33 for pions).
Figure 3 shows $r_{AA}$ for pions which has a suppression starting at $y_t$
= 3.0 ($p_t$ = 1.5 GeV/c). This shift between suppression of charged particle
at $p_t$ $>$ 2.0 GeV/c and suppression of pions at $p_t$ = 1.5 GeV/c is made up
by an enhancement of baryons\cite{Adler}. Proton $r_{AA}$ (see Figure 12)
which is the most numerous baryon shows an enhancement starting at 
$p_t$ $>$ 1.5. This enhancement continues to $p_t$ = 4.0 
($y_t$ = 4.0)\footnote{Note $y_t$ is calculated using a pion mass.}. Finally
at $p_t$=10.0 GeV/c ($y_t$ = 5.0) $r_{AA}$ = 0.2 which is the same value as 
pions.

\begin{figure*}[ht] \centerline{\includegraphics[width=0.800\textwidth]
{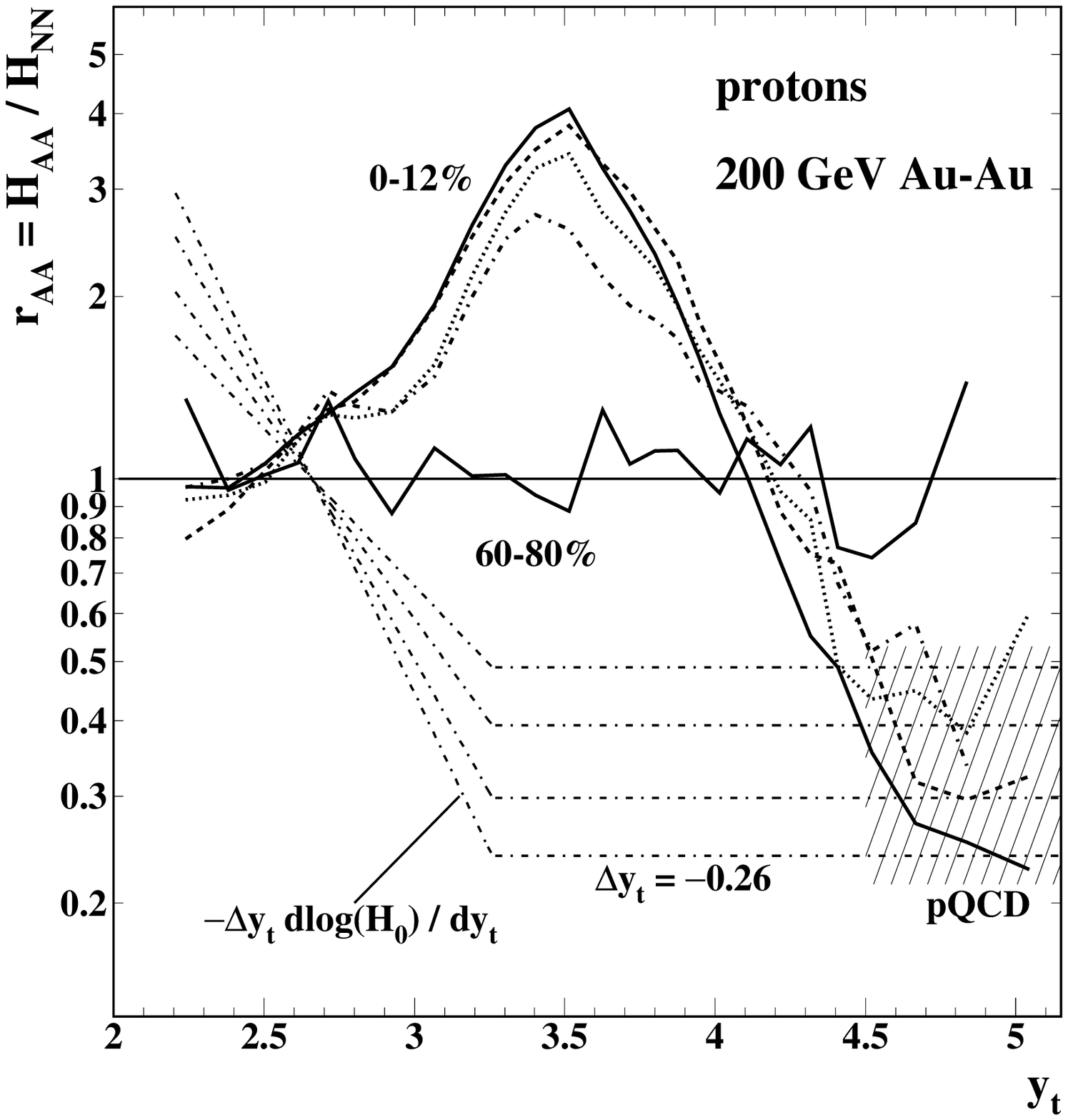}} \caption{Hard-component ratio for protons and five Au-Au 
centralities (thicker curves with changing line styles) relative to the 
N-N hard-component reference. The features are only comparable to Figure 3 for 
60-80\% or $y_t$ $>$ 4.5. Note $y_t$ is calculated using a pion mass.}
\label{fig11}
\end{figure*}

\subsection{Minijet Model} 

There is no \it a priori \rm model for the proton hard component. The excess
in the proton hard component \it appears \rm anomalous, but may be simply
explained in terms of parton energy loss\cite{Trainor2}.

\subsection{Glasma Flux Tube Model} 

We have shown previously that inside the tube there are three to four partons
with differing longitudinal momenta all at the same $\phi$. The $p_t$
distribution of the partons inside the tube is similar to pQCD but has a
suppression in the high $p_t$ region like the data. The partons which are
mainly gluons shower into more gluons which then fragment into quarks and 
anti-quarks which overlap in space and time with other quarks and anti-quarks 
from other partons. This leads to an enhance possibility for pairs of 
quarks from two different fragmenting partons to form a di-quark, since the 
recombining partons are localized together in a small volume. The same process 
will happen for pairs of anti-quarks forming a di-anti-quark. This 
recombination process becomes an important possibility in the GFTM compared 
to regular jet fragmentation. Since the quarks which overlap have similar 
phase space, the momentum of the di-quark is approximately twice the 
momentum of the quarks, but has approximately the same velocity. When
mesons are formed quarks pick up anti-quarks with similar phase space 
from fragmenting gluons to form a color singlet state. Thus the meson has
approximately twice the momentum of the quark and anti-quark of which it is
made. When the di-quark picks up a quark and forms a color singlet it will have
approximately three times the momentum of one of the three quarks it is made
from. Thus we expect a $p_t$ spectrum scaling when we compare mesons to
baryons. Figure 12 shows the ratio protons plus anti-protons to charged
particles as a function of $p_t$ for particles in our simulated central Au-Au
collisions. In Figure 12 we also plot the ratio from central Au-Au RHIC
data\cite{Adler}. These experimental results agree well (considering the
errors) with the GFTM predictions for all charged particles. The
background particles which came from HIJING\cite{hijing} have the same ratios 
observed in p-p collisions, while particles coming from our tube have a much 
larger ratio.

\begin{figure*}[ht] \centerline{\includegraphics[width=0.800\textwidth]
{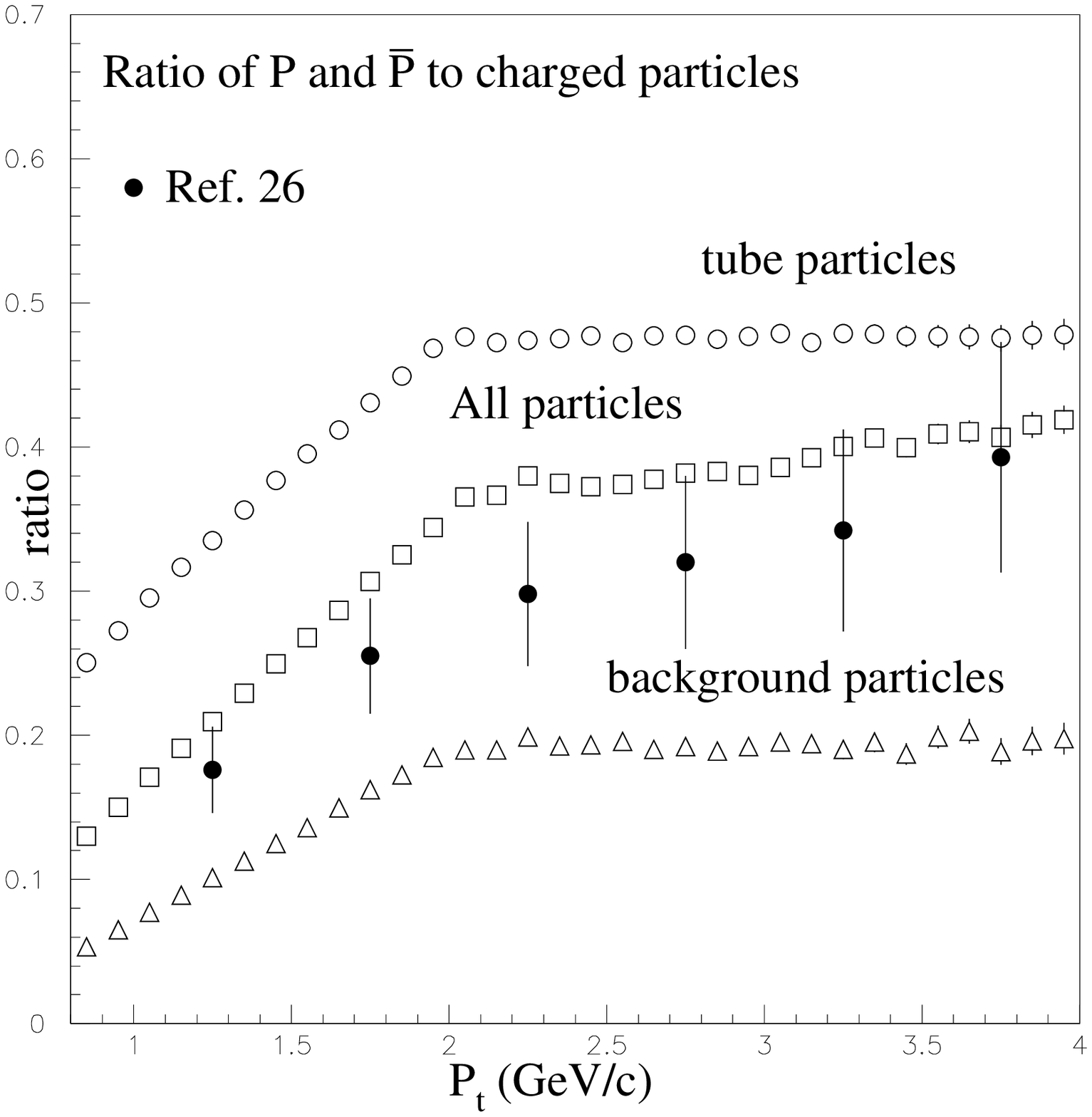}} \caption{Shows the ratio of protons plus anti-protons to 
charged particles as a function of $p_t$ for particles in our simulated 
central Au-Au collisions. We also plot the ratio from central Au-Au RHIC data. 
These experimental results agree well (considering the errors) with the GFTM 
predictions for all charged particles. The plotted ratio for the background 
particles coming from HIJING is similar to p-p collisions.}
\label{fig12}
\end{figure*}

\section{The Ridge is formed by the Flux Tubes when a jet trigger is added to 
the GFTM}

In heavy ion collisions at RHIC there has been observed a phenomenon called the
ridge which has many different explanations[28-35]. The ridge is a long
range charged particle correlation in $\Delta \eta$ (very flat), while the 
$\Delta \phi$ correlation is approximately jet-like (a narrow Gaussian). 
There also appears with the ridge a jet-like charged-particle-pair correlation 
which is symmetric in $\Delta \eta$ and $\Delta \phi$ such that the peak of the
jet-like correlation is at $\Delta \eta$ = 0 and $\Delta \phi$ = 0. The 
$\Delta \phi$ correlation of the jet and the ridge are approximately the same 
and smoothly blend into each other. The ridge correlation is generated when 
one triggers on an intermediate $p_t$ range charged particle and then forms
pairs between that trigger particle and each of all other intermediate charged 
particles with a smaller $p_t$ down to some lower limit. 

\subsection{STAR experiment measurement of the ridge}

Triggered angular correlation data showing the ridge was presented at Quark 
Matter 2006\cite{Putschke}. Figure 14 shows the experimental $\Delta \phi$ vs. 
$\Delta \eta$ CI correlation for the 0-10\% central Au-Au collisions at 
$\sqrt{s_{NN}} =$ 200 GeV requiring one trigger particle $p_t$ between 3.0 and 
4.0 GeV/c and an associated particle $p_t$ above 2.0 GeV/c. The yield is
corrected for the finite $\Delta \eta$ pair acceptance. 

\begin{figure*}[ht] \centerline{\includegraphics[width=0.800\textwidth]
{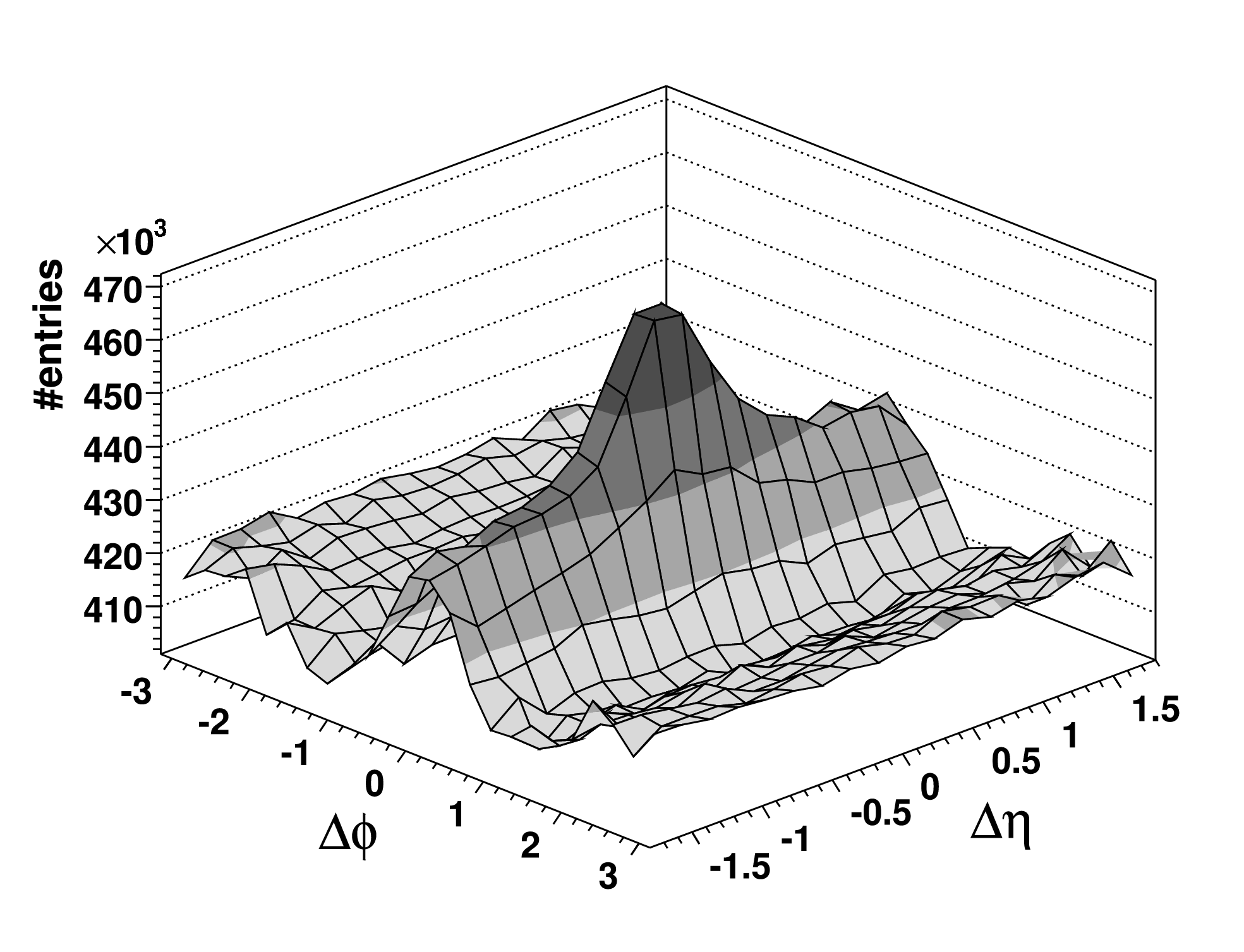}} \caption{Raw $\Delta \phi$ vs. $\Delta \eta$ CI preliminary
correlation data\cite{Putschke} for the 0-10\% centrality bin for Au-Au
collisions at $\sqrt{s_{NN}} =$ 200 GeV requiring one trigger particle $p_t$
between 3.0 to 4.0 GeV/c and an associated particle $p_t$ above 2.0 GeV/c.}
\label{fig13}
\end{figure*}

In this paper we will investigate whether the GFTM can account for the ridge 
once we add a jet trigger to GFTM generator\cite{PBMtoGFTM}. However this 
trigger will also select jets which previously could be neglected because 
there was such strong quenching[37-39] of jets in central collisions when a 
jet trigger had not been used. We use HIJING\cite{hijing} to determine the 
number of expected jets. We have already shown that our final state particles 
come from hadrons at or near the fireball surface. We reduce the number of jets
by 80\% which corresponds to the estimate that only the parton interactions on 
or near the surface are not quenched away, and thus at kinetic freezeout 
fragment into hadrons which enter the detector. This 80\% reduction is 
consistent with single $\pi^0$ suppression observed in Ref.\cite{quench3}. 
We find for the reduced HIJING jets that 4\% of the Au-Au central events 
(0-5\%) centrality at $\sqrt{s_{NN}} =$ 200 have a charged particle with a 
$p_t$ between 3.0 and 4.0 GeV/c with at least one other charged particle 
with its $p_t$ greater than 2.0 GeV/c coming from the same jet. With the 
addition of the quenched jets to the generator, we then form 
two-charged-particle correlations between one-charged-particle with a $p_t$ 
between 3.0 to 4.0 GeV/c and another charged-particle whose $p_t$ is greater 
than 2.0 GeV/c. The results of these correlations are shown in Figure 15.

\begin{figure*}[ht] \centerline{\includegraphics[width=0.800\textwidth]
{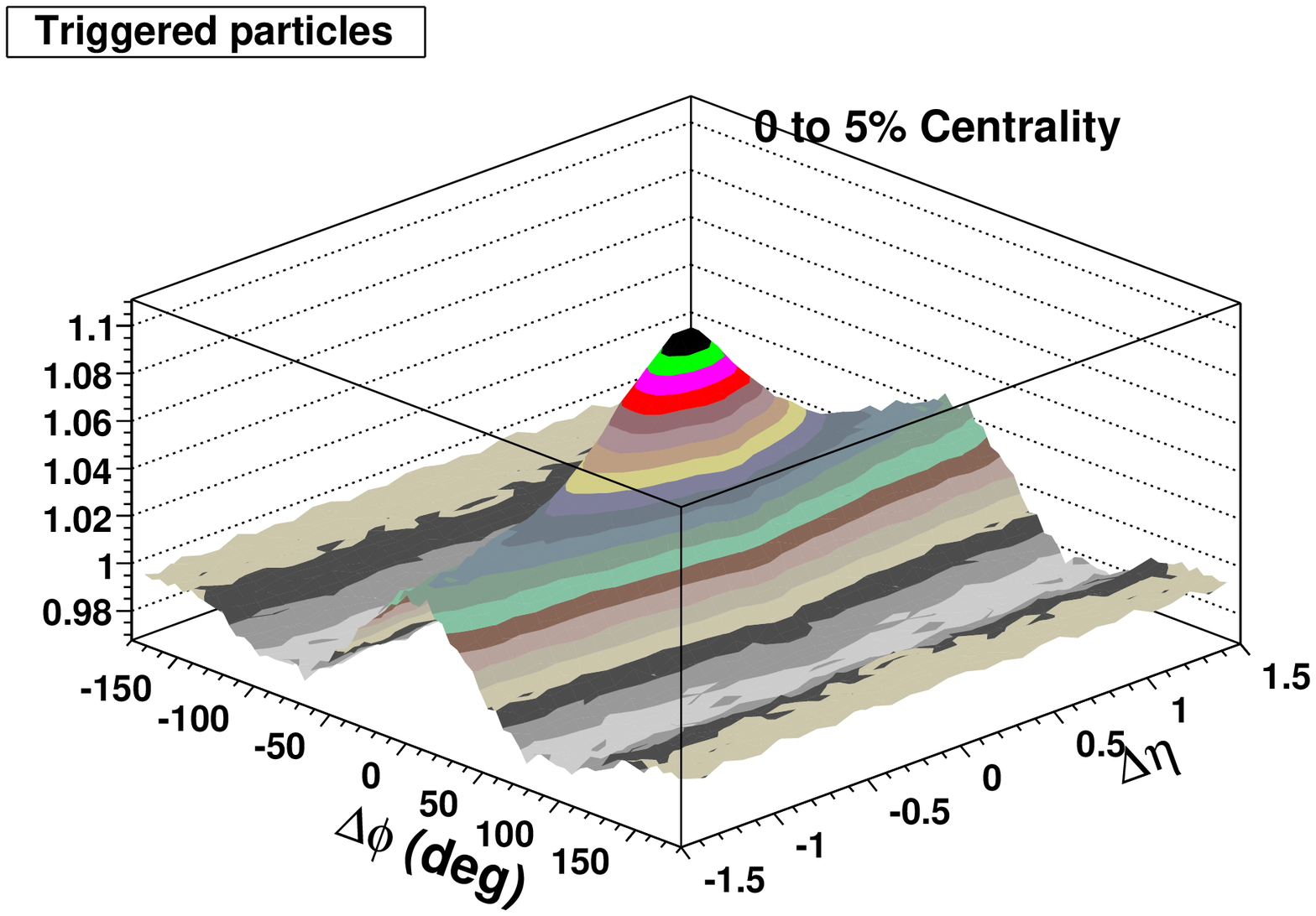}} \caption{The GFTM generated CI correlation for the 0-5\% 
centrality bin requiring one trigger particle $p_t$ above 3.0 GeV/c and less 
than 4.0 GeV/c and another particle $p_t$ above 2.0 GeV/c plotted as a two 
dimensional $\Delta \phi$ vs. $\Delta \eta$ perspective plot. The trigger 
requirements on this figure are the same as those on the experimental data in 
Figure 13.}
\label{fig14}
\end{figure*}

We can compare the two figures, if we realize that the away-side ridge has 
around 420,000 pairs in Figure 14 while in Figure 14 the away-side ridge has a 
correlation of around 0.995. If we multiply the correlation scale of Figure 14
by 422,111 in order to achieve the number of pairs seen in Figure 13, the 
away-side ridge would be at 420,000 and the peak would be at 465,000. This 
would make a good agreement between the two figures. We know in our Monte Carlo
which particles come from tubes which would be particles form the ridge. The 
correlation formed by the ridge particles is generated almost entirely by 
particles emitted by the same tube. Thus we can predict the shape and the yield
of the ridge for the above $p_t$ trigger selection and lower cut, by plotting 
only the correlation coming from pairs of particles that are emitted by the 
same tube (see Figure 15).

In Ref.\cite{Putschke} it was assumed that the ridge yield was flat across 
the acceptance while in Figure 14 we see that this is not the case. 
Therefore our ridge yield is 35\% larger than estimated in Ref.\cite{Putschke}.
Finally we can plot the jet yield that we had put into our Monte Carlo. 
The jet yield is plotted in Figure 16 where we subtracted the tubes and 
the background particles from Figure 14.

\begin{figure*}[ht] \centerline{\includegraphics[width=0.800\textwidth]
{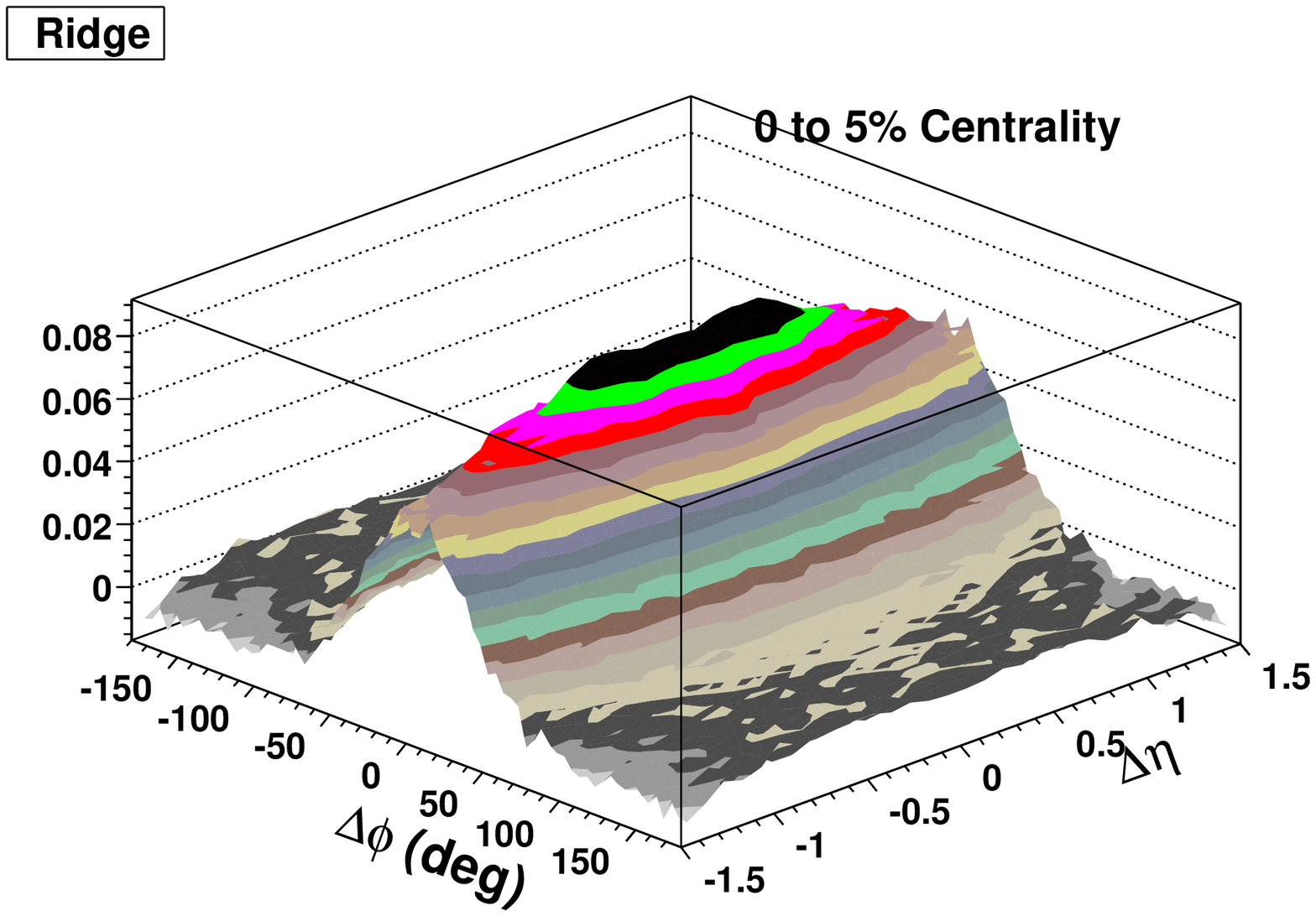}} \caption{The ridge signal is the piece of the CI correlation
for the 0-5\% centrality of Figure 15 after removing all other particle pairs
except the pairs that come from the same tube. It is plotted as a two
dimensional $\Delta \phi$ vs. $\Delta \eta$ perspective plot.}
\label{fig15}
\end{figure*}

\begin{figure*}[ht] \centerline{\includegraphics[width=0.800\textwidth]
{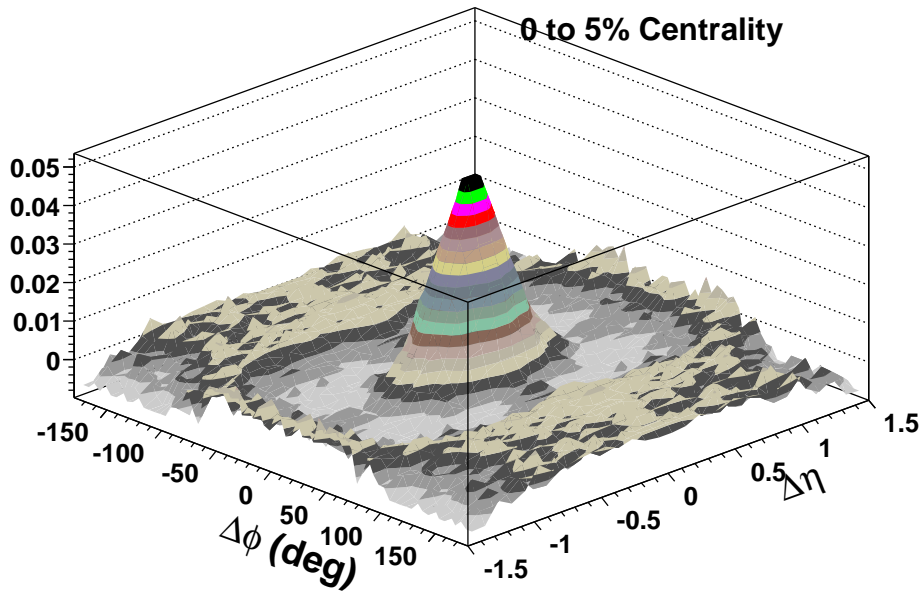}} \caption{The jet signal is left in the CI correlation after 
the contributions from the background and all the bubble particles are removed 
from the 0-5\% centrality (with trigger requirements) of Figure 15. It is 
plotted as a two dimensional $\Delta \phi$ vs. $\Delta \eta$ perspective plot.}
\label{fig16}
\end{figure*}

\clearpage

\subsection{PHOBOS experiment measurement of the ridge}

The PHOBOS detector select triggered charged tracks with $p_t$ $>$ 2.5GeV/c
within an acceptance of 0 $<$ $\eta^{trig}$ $<$ 1.5. Associated charge
particles that escape the beam pipe are detected in a range $|\eta|$ $<$ 
3.0. The CI correlation is shown in Figure 17\cite{PHOBOS}. The near-side
structure is more closely examined by integrating over $|\Delta \phi|$ $<$ 1
and is plotted in Figure 18 as a function of $\Delta \eta$. PYTHIA simulation
for p-p events is also shown. Bands around the data points represent the 
uncertainty from flow subtraction. The error on the ZYAM procedure is shown 
as a gray band at zero. All systematics uncertainties are 90\% confidence 
level.
 
\begin{figure*}[ht] \centerline{\includegraphics[width=0.800\textwidth]
{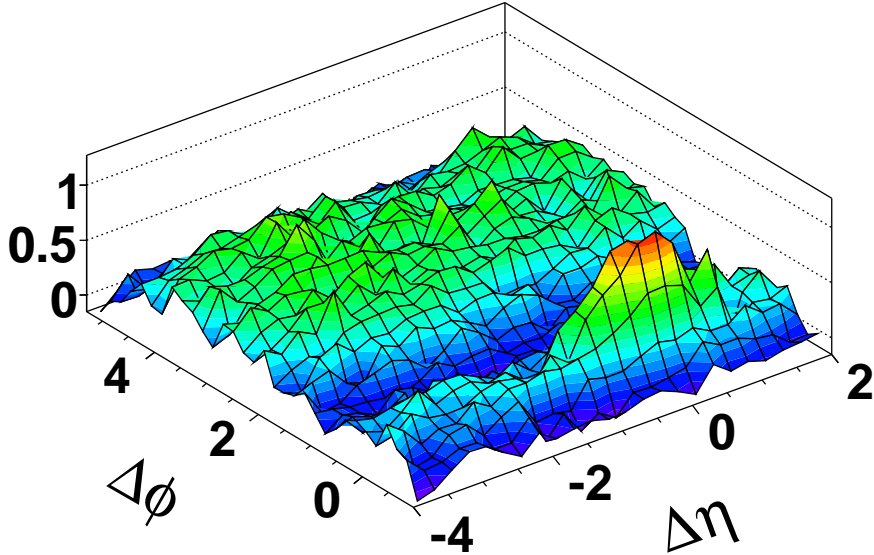}} \caption{PHOBOS triggered CI correlation from 0-30\% central 
Au-Au collisions. It is plotted as a two dimensional $\Delta \phi$ vs. 
$\Delta \eta$ perspective plot.}
\label{fig17}
\end{figure*}

\begin{figure*}[ht] \centerline{\includegraphics[width=0.800\textwidth]
{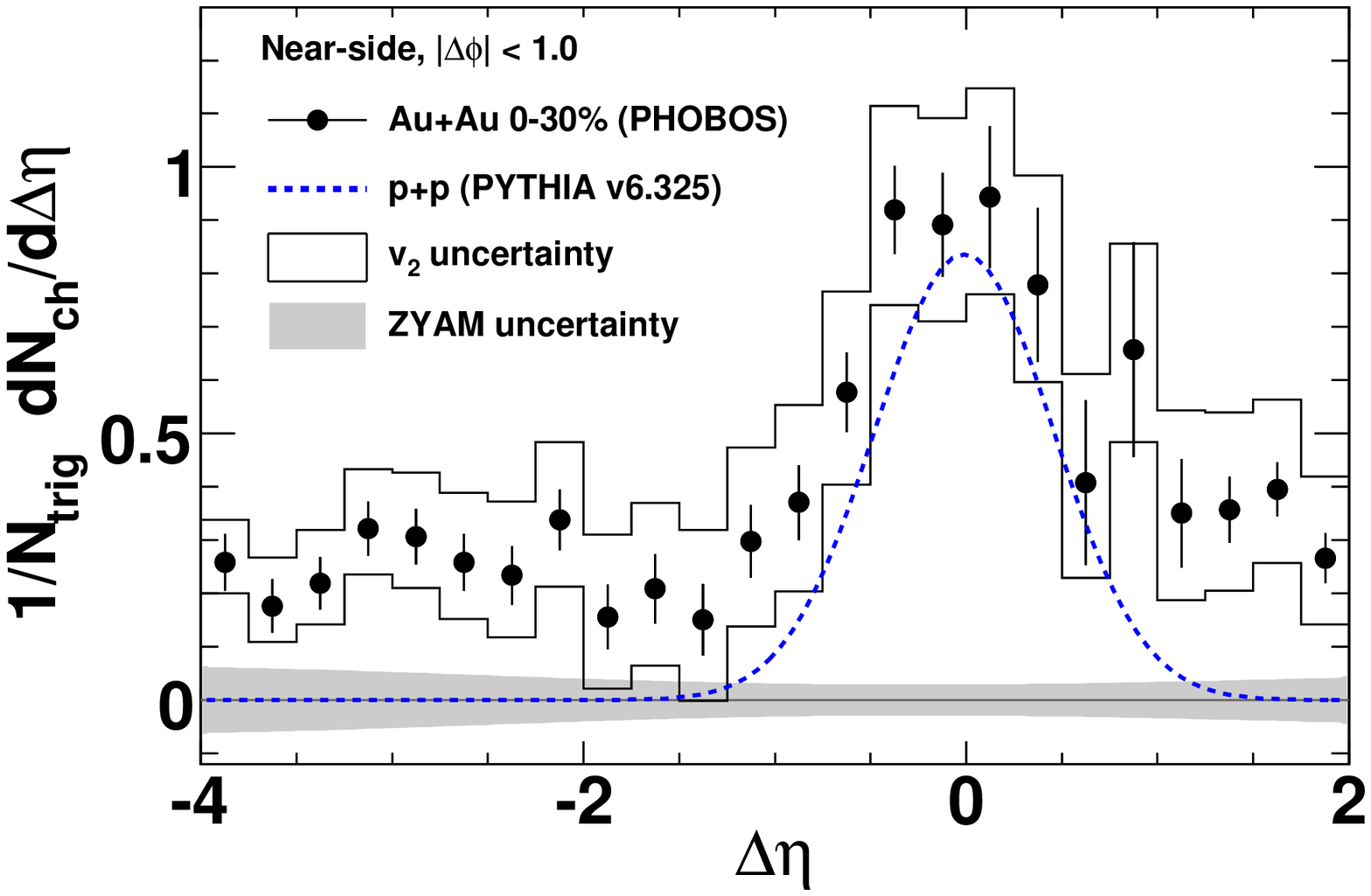}} \caption{PHOBOS near-side trigger yield integrated over 
$|\Delta \phi|$ $<$ 1.0 for 0-30\% central Au-Au compared to PYTHIA p-p 
(dashed line) as a function of $\Delta \eta$. Bands around the data points 
represent the uncertainty from flow subtraction. The error on the ZYAM 
procedure is shown as a gray band at zero. All systematics uncertainties are 
90\% confidence level.}
\label{fig18}
\end{figure*}

We generate using our above GFTM the two-charged-particle correlations between 
one-charged-particle with a $p_t$ between greater than 2.5 GeV/c which has an 
$|\eta|$ $<$ 0.75 and another charged-particle whose $p_t$ is greater than 7 
MeV/c in a range $|\eta|$ $<$ 3.0. The results of this correlation is shown in 
Figure 20. We see that the triggered correlation is vary similar to the PHOBOS
results. In order to make a comparison we integrate the near-side structure 
over $|\Delta \phi|$ $<$ 1 and plot it as a smooth curve in Figure 21 as a 
function of $\Delta \eta$. We also again plot the PHOBOS points on the same 
plot. The long range correlation over $\Delta \eta$ produced by the GFTM is 
possible and does not violate causality, since the glasma flux tubes are 
generated early in the collision. The radial flow which develops at a later 
time pushes the surface tubes outward in the same $\phi$ direction because the 
flow is purely radial. In order to achieve such an effect in minijet 
fragmentation one would have to have fragmentation moving faster than the speed
of light.

\begin{figure*}[ht] \centerline{\includegraphics[width=0.800\textwidth]
{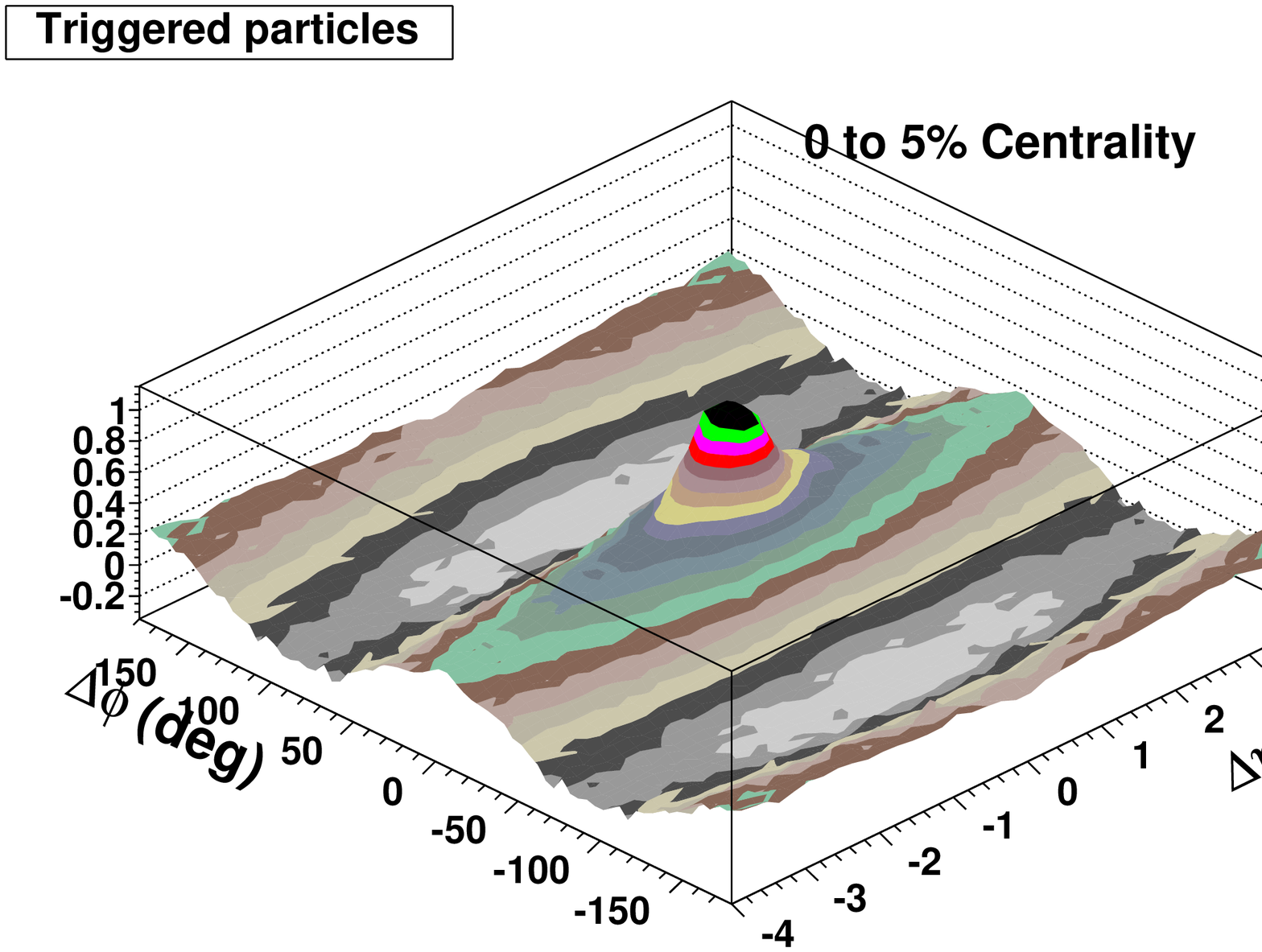}} \caption{GFTM triggered (see text) CI correlation from 0-5\% 
central Au-Au collisions. It is plotted as a two dimensional $\Delta \phi$ vs. 
$\Delta \eta$ perspective plot which is the same as Figure 18.}
\label{fig19}
\end{figure*}

\begin{figure*}[ht] \centerline{\includegraphics[width=0.800\textwidth]
{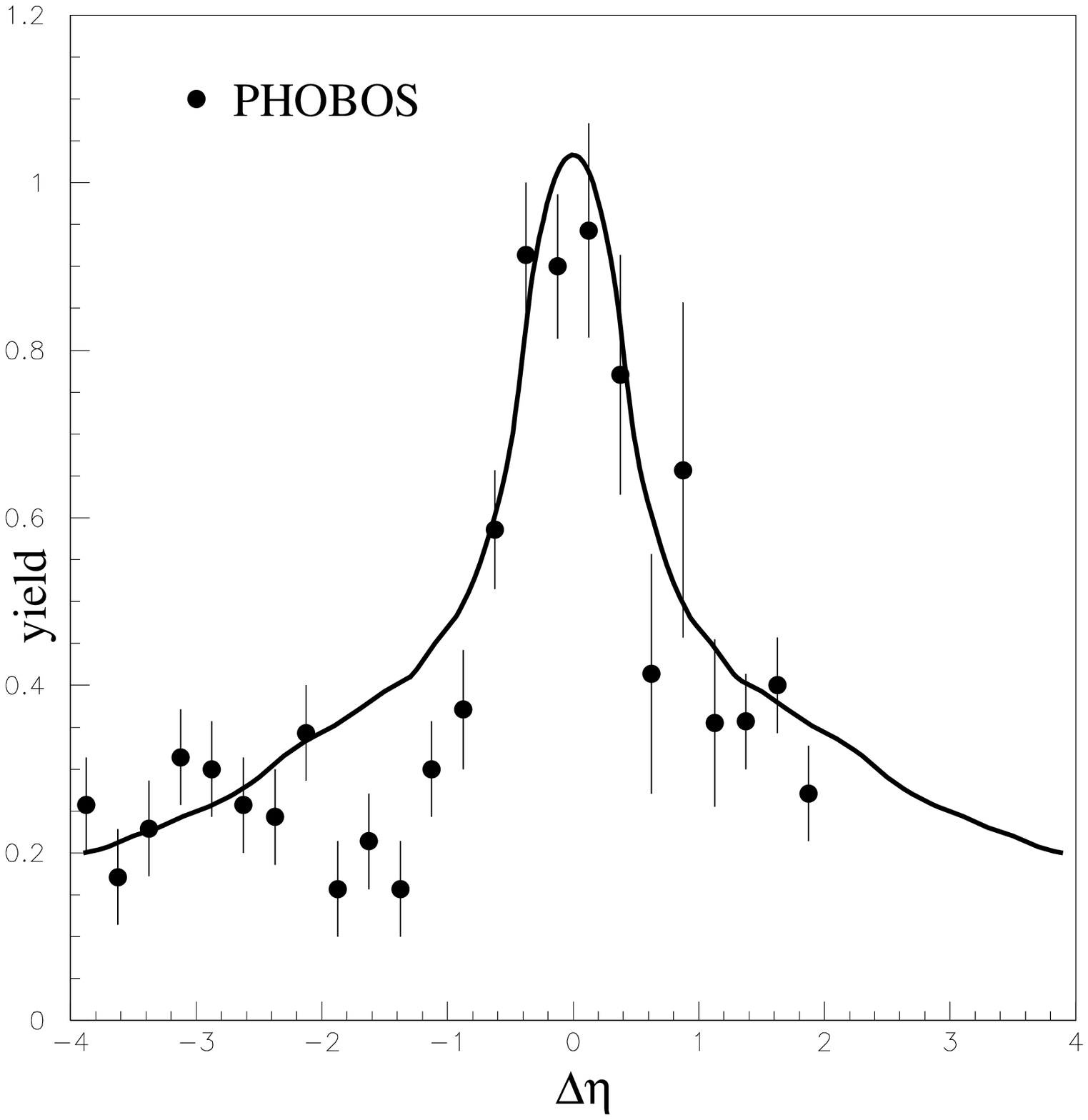}} \caption{GFTM $\phi$ integrated ($|\Delta \phi|$ $<$ 1) 
near-side structure compared to PHOBOS near-side structure (see Figure 20).} 
\label{fig20}
\end{figure*}

\section{The Mach Cone Effect and Three Particle Angular Correlations}

It was reported by the PHENIX Collaboration\cite{PHENIXCONE} that the shape
of the away-side $\Delta \phi$ distributions for non-peripheral collisions 
are not consistent with purely stochastic broadening of the peripheral
Au-Au away-side. The broadening and possible change in shape of the away-side
jet are suggestive of a mach cone\cite{renk}. The mach cone effect depended
on the trigger particles and the associated particles used in the two particle
angular correlations. The mach cone shape is not present if one triggers
on 5-10 GeV/c particles and also use hard associated particles greater than 3 
GeV/c\cite{jiangyong}. The away-side broadening in the two-particle depends on
the $v_2$ subtraction and the method of smearing the away-side dijet component
and momentum conservation.

One needs to go beyond two particle correlations in order to learn more. A
three particle azimuthal angle correlation should reveal very clear pattern
for the mach cone compared to other two particle method. The STAR 
Collaboration\cite{STARCONE} has made such correlation studies and does find 
a structure some what like a mach cone. However the trigger dependence as seen 
in Ref.\cite{jiangyong} and the measurement of the conical emission angle to 
be independent of the associated particle $p_t$ in Ref.\cite{STARCONE} is not 
consistent with a mach cone. The similarity of the mach cone and the ridge is 
very interesting to consider and makes one consider that they are the same 
effect. In the last section we showed that flux tubes can explain the ridge, 
thus they should explain the mach cone. One would not expect the minijet model 
will give a mach cone like correlation.

We saw at the end of Sec. 2 that final state anisotropic flow pattern can be 
decomposed in a Fourier series ($v_1$, $v_2$, $v_3$, ...). When one triggers
on a particle coming from a flux tube, the other flux tubes contribute the 
components of the anisotropic flow pattern. In Figure 22 we show two such tube
arrangements. In order to make contact with the data and show the difference
between the minijet model and GFTM we will define a trigger particle and a
reference particle(s). We choose a $p_t$ of greater than 1.1 GeV/c for the
trigger and less than this value for the reference. The two particle 
correlation for this trigger and reference for the minijet model and the GFTM
are equal to each other.

\begin{figure*}[ht] \centerline{\includegraphics[width=0.800\textwidth]
{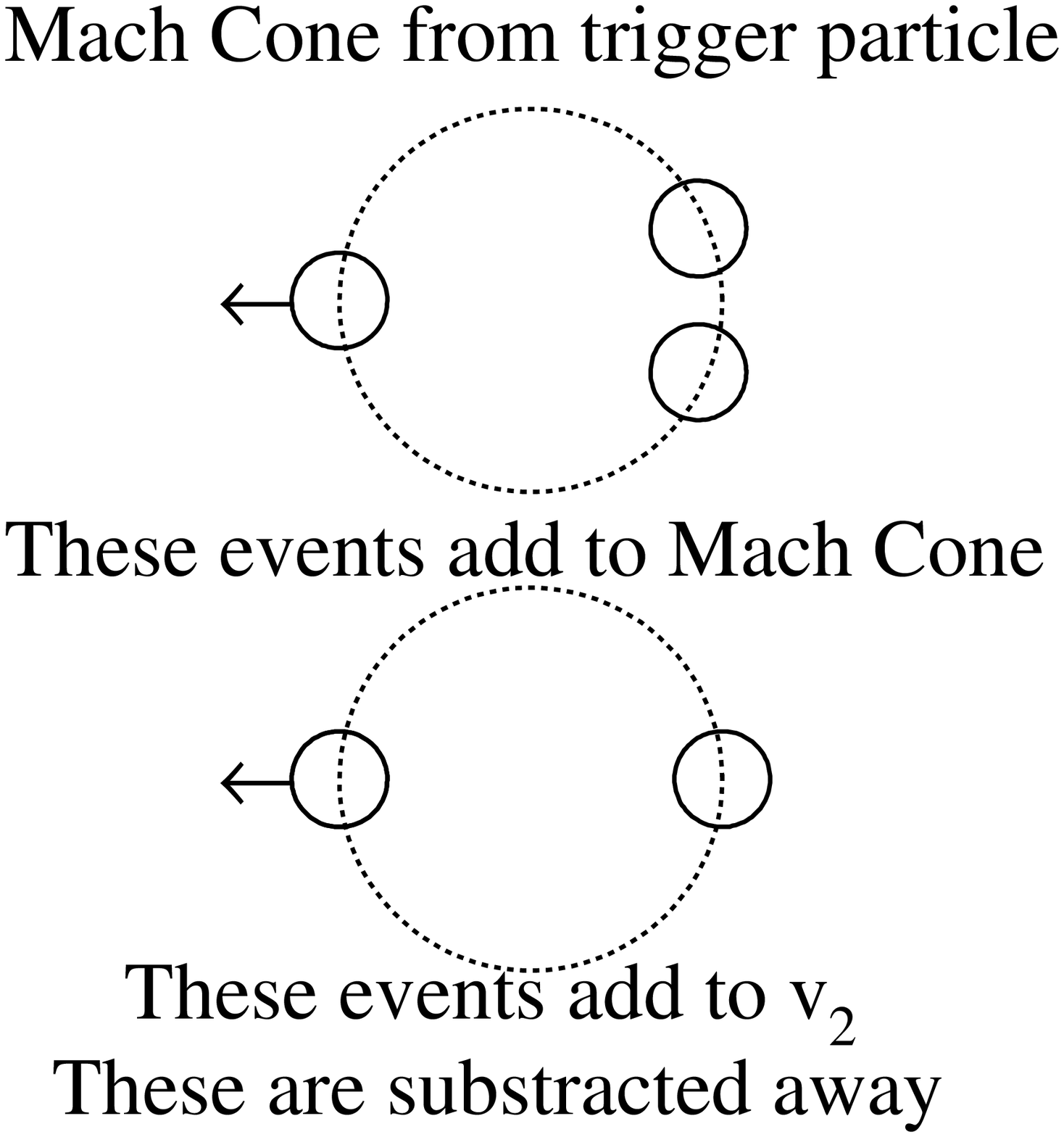}} \caption{Top arrangement has a $v_3$ component while the bottom
gives a $v_2$ component. In most mach cone analyses $v_2$ is subtracted.}
\label{fig21}
\end{figure*}

We define a three particle correlation using the azimuthal angle between a
trigger particle 1 and a reference particle 2 ($\Delta \phi_{12}$) and for the 
third particle the azimuthal angle between a trigger particle 1 and a 
reference particle 3($\Delta \phi_{13}$). The two particle correlation of the 
two models are the same and the three particle correlation is nearly the same. 
In order to obtain the true three particle effect we must remove the two 
particle correlation from the raw three particle correlation. This removal 
gives the so-called three particle cumulant. Figure 22 and Figure 23 shows the 
three particle cumulant for $\Delta \phi_{12}$ vs. $\Delta \phi_{13}$ for the 
minijet model and the GFTM. 

\begin{figure*}[ht] \centerline{\includegraphics[width=0.800\textwidth]
{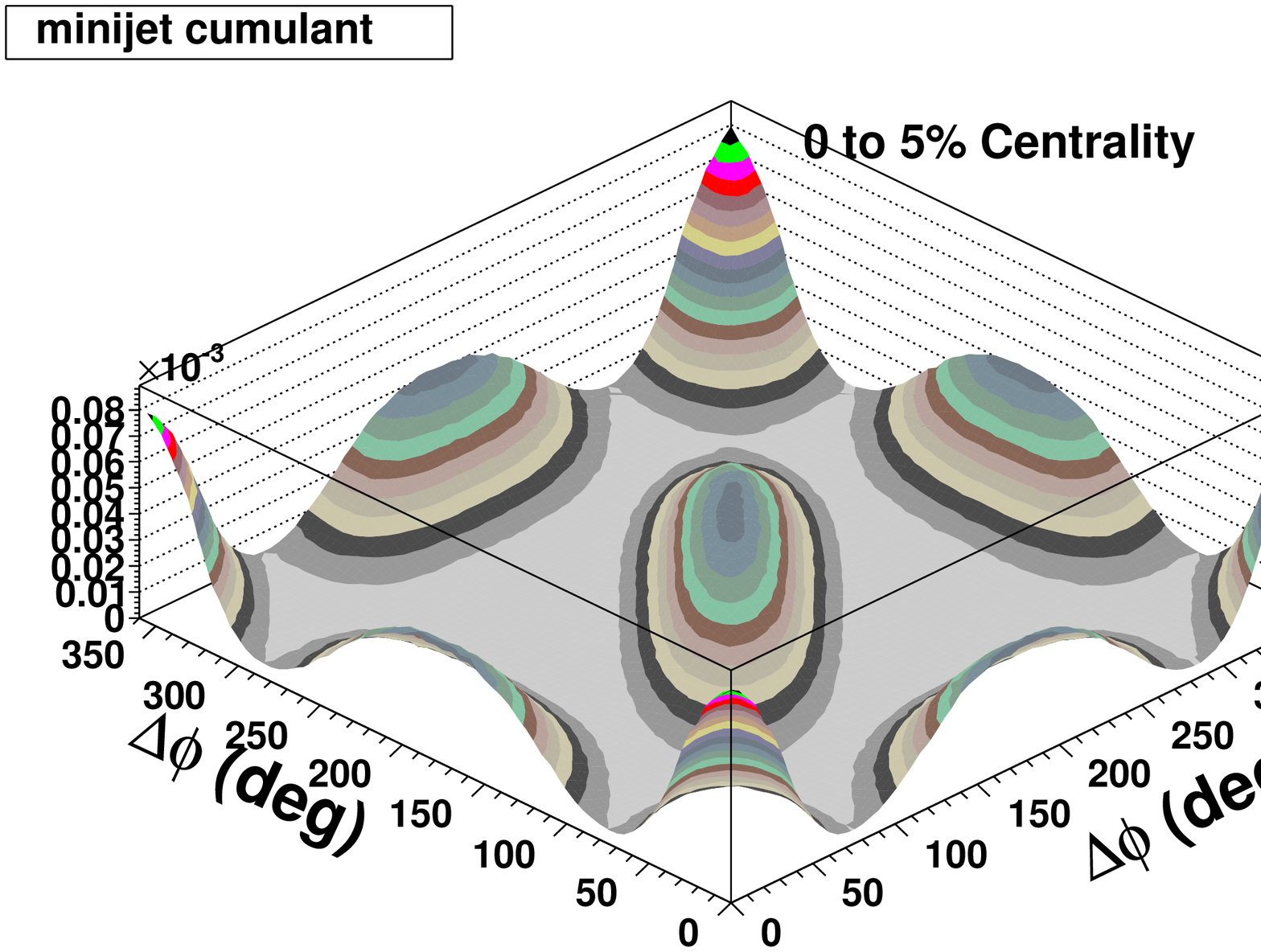}} \caption{The minijet generated three particle cumulant for 
the 0-5\% centrality bin requiring one trigger particle $p_t$ above 1.1 GeV/c 
and other reference particles $p_t$ below 1.1 GeV/c plotted as a two 
dimensional combinations of trigger particle 1 paired with two reference 
particles 2 and 3 creating $\Delta \phi_{12}$ vs. $\Delta \phi_{13}$ 
perspective plot.}
\label{fig22}
\end{figure*}

\begin{figure*}[ht] \centerline{\includegraphics[width=0.800\textwidth]
{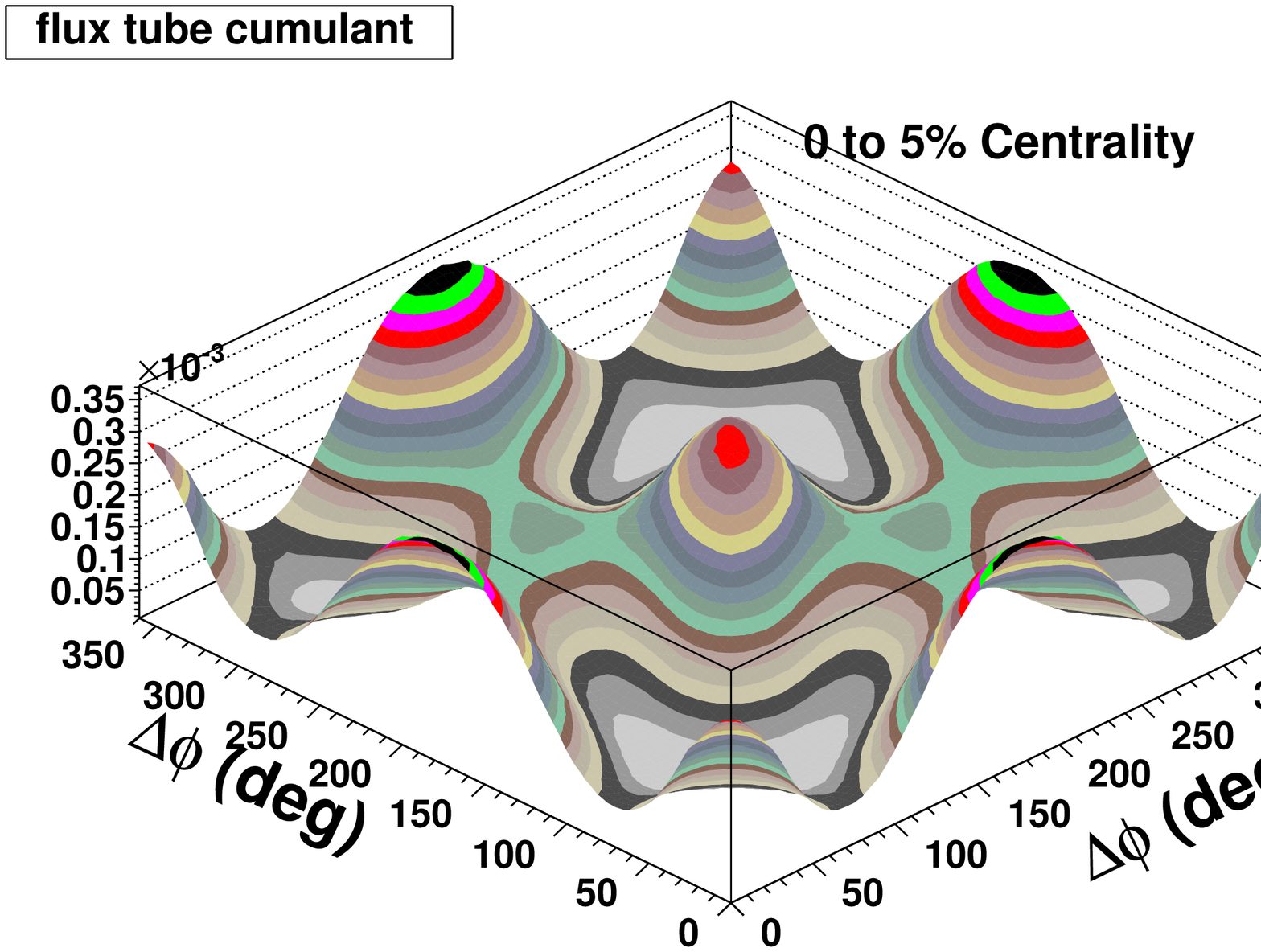}} \caption{The GFTM generated three particle cumulant for the 
0-5\% centrality bin requiring one trigger particle $p_t$ above 1.1 GeV/c and 
other reference particles $p_t$ below 1.1 GeV/c plotted as a two dimensional 
combinations of trigger particle 1 paired with two reference particles 2 and 3 
creating $\Delta \phi_{12}$ vs. $\Delta \phi_{13}$ perspective plot.}
\label{fig23}
\end{figure*}

The minijet model Figure 22 shows only diagonal away-side response coming from 
the underlying dijet nature of the minijets. The GFTM Figure 23 also shows the
diagonal away-side response of momentum conservation, however $v_3$ 
configurations of Figure 21 gives an off-diagonal island at $\Delta \phi_{12}$
$\sim$ $130^\circ$,  $\Delta \phi_{13}$ $\sim$ $230^\circ$ or vice versa.  

Let us look at the reported STAR data\cite{Ulery} in order to address this 
off-diagonal effect. In Figure 24 we show the raw three particle results from 
STAR where (a) is the raw two particle correlation (points). They also show in
(a) the background formed from mixed events with flow modulation added-in 
(solid). The background subtracted two particle correlation is shown as an 
inset in (a) where the double hump mach cone effect is clear. In Figure 24(b) 
the STAR raw three particle correlation is shown which required one trigger 
particle 3.0 $<$ $p_t$ $<$ 4.0 GeV/c and other reference particles 2 and 3 
with 1.0 $<$ $p_t$ $<$ 2.0 GeV/c. An off-diagonal island does appear at 
$\Delta \phi_{12}$ $\sim$ 2.3,  $\Delta \phi_{13}$ $\sim$ 4.0 radians or 
vice versa. This is the same values as the island which occurs in the GFTM
and is the off-diagonal excess that is claimed to be the mach cone. Like
the ridge effect the mach cone seems to be the left overs of the initial 
state flux tube arrangements related to the fluctuations in the third harmonic.
Higher order harmonic fluctuations become less likely to survive to the final
state.

\begin{figure*}[ht] \centerline{\includegraphics[width=0.800\textwidth]
{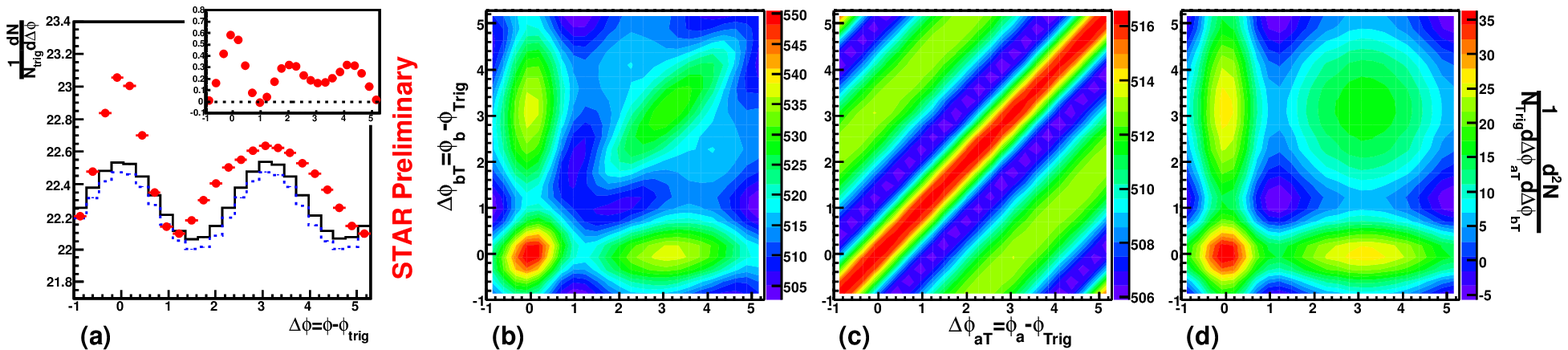}} \caption{(a) STAR data raw two particle correlation points, 
background from mixed events with flow modulation added-in (solid) along with 
the background subtracted two particle correlation (inset). (b) Raw three 
particle correlation from STAR requiring one trigger particle 3.0 $<$ $p_t$ 
$<$ 4.0 GeV/c and other reference particles 2 and 3 with 1.0 $<$ $p_t$ $<$ 2.0 
GeV/c creating a two dimensional perspective plot $\Delta \phi_{12}$ vs. 
$\Delta \phi_{13}$.}
\label{fig24}
\end{figure*}

\section{Summary and Discussion}

In this article we have made a comparison between two very different models
for central Au-Au collisions. Both models are successful at describing the
the spectrum of particles produced and the two particle angular correlations
observed in ultrarelativistic heavy ion collisions. The first model is a 
minijet model which assumes that around $\sim$50 minijets are produced in 
back-to-back pairs and has an altered fragmentation function from that of
vacuum fragmentation. It also assumes that the fragments are transparent and 
escape the collision zone and are then detected. The second model is a glasma 
flux tube model which leads to longitudinal color electric and magnetic fields
confined in flux tubes on the surface of a radial expanding fireball driven 
by interacting flux tubes near the center of the fireball through plasma 
instabilities. This internal fireball becomes an opaque hydro fluid which 
pushes the surface flux tubes outward. Around $\sim$12 surface flux tubes 
remain and fragment with $\sim$1/2 the produced particles escaping the 
collision zone and are detected. 

We expand our comparison to other phenomenon of the central collisions. We
considered in Sec. 3 baryon and anti-baryon formation in both models. 
There was no \it a priori \rm reason for the excess in the minijet model,
while in the glasma flux tube model(GFTM) recombination of quarks into 
di-quarks and anti-quark into anti-di-quarks leads to a natural excess of
baryon and anti-baryon formation in this model. 

The formation of the ridge phenomena is discussed in Sec. 4. In order to 
achieve a long range correlation effect in minijet fragmentation one would 
have to have fragmentation moving faster than the speed of light. The
GFTM however can have a long range correlation over $\Delta \eta$ since the 
glasma flux tubes are generated early in the collision. The radial flow which 
develops at a later time pushes the surface tubes outward in the same 
$\phi$ direction because the flow is purely radial. Thus a long range 
$\Delta \eta$ last to the final state of the collision. A very good comparison
was achieved between data and GFTM. 

Sec. 5 treats the so-called mach cone effect by analyzing three particle 
angular correlations in the two models. The minijet model and the GFTM
have the same two particle angular correlations but when the three particle
azimuthal angular correlations are compared the two models differ. The minijet 
model shows only diagonal away-side response coming from the underlying dijet 
nature of the minijets, while GFTM also shows a diagonal away-side response it
also shows an off-diagonal island. Like the ridge effect the mach cone seems 
to be left over from the initial state flux tube arrangements related to the 
fluctuations in the third harmonic($v_3$). This off-diagonal island excess is
seen in the data and is claimed to be the mach cone.

Relativistic Heavy Ion Collider (RHIC) collisions are conventionally described 
in terms of two major themes: hydrodynamic evolution of a thermal bulk medium 
and energy loss of energetic partons in that medium through gluon 
bremsstrahlung. The minijet model is not consistent with is standard view.
The glasma flux tube model generates a fireball which becomes an opaque
hydro fluid that is consistent with conventional ideas. Even though both
models can obtain the same spectrum of particles and the same two particle 
angular correlations, it is only the GFTM that can tie all together.

\section{Acknowledgments}

This research was supported by the U.S. Department of Energy under Contract No.
DE-AC02-98CH10886. The author thank Sam Lindenbaum and William Love for 
valuable discussion and Bill for assistance in production of figures. It is 
sad that both are now gone.

\end{document}